\documentclass[fleqn,usenatbib]{mnras}

\usepackage{newtxtext,newtxmath}

\usepackage[T1]{fontenc}

\DeclareRobustCommand{\VAN}[3]{#2}
\let\VANthebibliography\thebibliography
\def\thebibliography{\DeclareRobustCommand{\VAN}[3]{##3}\VANthebibliography}

\usepackage{graphicx}	
\usepackage{amsmath}	

\title[Stellar wind impact on terrestrial early atmospheres]{Stellar wind impact on early atmospheres around unmagnetized Earth-like planets}

\author[A. Canet et al.]{
Ada Canet,$^{1,2,3}$\thanks{E-mail: adacanet@ucm.es (AC)}
Jacobo Varela,$^{4,5}$
and Ana I. Gómez de Castro$^{1,2}$
\\
$^{1}$Joint Center for Ultraviolet Astronomy (JCUVA), Universidad Complutense de Madrid, Madrid, Spain\\
$^{2}$Facultad de Ciencias Matemáticas, S.D. Astronomia y Geodesia, Universidad Complutense de Madrid, Madrid, Spain\\
$^{3}$Facultad de Ciencias Físicas, Departamento de Física de la Tierra y Astrofísica, Universidad Complutense de Madrid, Madrid, Spain\\
$^{4}$Universidad Carlos III de Madrid, Leganes, 28911, Spain\\
$^{5}$Institute for Fusion Studies, Department of Physics, University of Texas at Austin, Austin, Texas 78712, USA
}

\date{Accepted XXX. Received YYY; in original form ZZZ}

\pubyear{2024}

\begin{document}
\label{firstpage}
\pagerange{\pageref{firstpage}--\pageref{lastpage}}
\maketitle

\begin{abstract}
Stellar rotation at early ages plays a crucial role in the survival of primordial atmospheres around Earth-mass exoplanets. Earth-like planets orbiting fast-rotating stars may undergo complete photoevaporation within the first few hundred Myr driven by the enhanced stellar XUV radiation, while planets orbiting slow-rotating stars are expected to experience difficulty to lose their primordial envelopes. Besides the action of stellar radiation, stellar winds induce additional erosion on these primordial atmospheres, altering their morphology, extent, and causing supplementary atmospheric losses. In this paper, we study the impact of activity-dependent stellar winds on primordial atmospheres to evaluate the extent at which the action of these winds can be significant in the whole planetary evolution at early evolutionary stages. We performed 3D magnetohydrodynamical (MHD) simulations of the interaction of photoevaporating atmospheres around unmagnetized Earth-mass planets in the time-span between 50 and 500 Myr, analyzing the joint evolution of stellar winds and  atmospheres for both fast- and slow-rotating stars. Our results reveal substantial changes in the evolution of primordial atmospheres when influenced by fast-rotating stars, with a significant reduction in extent at early ages. In contrast, atmospheres embedded in the stellar winds from slow-rotating stars remain largely unaltered. The interaction of the magnetized stellar winds with the ionized upper atmospheres of these planets allows to evaluate the formation and evolution of different MHD structures, such as double-bow shocks and induced magnetospheres. This work will shed light to the first evolutionary stages of Earth-like exoplanets, that are of crucial relevance in terms of planet habitability.
\end{abstract}




\section{Introduction}
\defcitealias{2015ApJ...815L..12J}{J15}

During the first stages of their evolution, protoplanetary cores with masses above $\sim 0.1 M_{\oplus}$ are able to accrete a significant H/He-rich atmosphere directly from the protoplanetary disk \citep{2006ApJ...648..696I,2015ApJ...811...41L,2016ApJ...825...86S,2020SSRv..216..129O}. Retention of these primordial atmospheres depends on the mass of the planet that gravitationally bounds the gas, and the X-ray and extreme ultraviolet (EUV) radiation of the star that is absorbed in the upper atmosphere, heating the atmospheric gas, and leading to thermal hydrodynamic escape \citep{2013AsBio..13.1011E,2014MNRAS.439.3225L,2015ApJ...815L..12J,2018A&A...619A.151K}. 

The XUV (EUV+X-ray) radiation is a byproduct of the stellar magnetic activity, controlled by the rotation-powered stellar dynamo, with rapidly rotating stars showing an enhanced XUV emission \citep{2003A&A...397..147P,2011ApJ...743...48W,2014ApJ...794..144R}. Thus, the stellar magnetic activity plays a crucial role in shaping the evolution of initial hydrogen-rich atmospheres. \citet[][hereafter \citetalias{2015ApJ...815L..12J}]{2015ApJ...815L..12J} modeled the evolution of Earth-mass primordial H-rich atmospheres with the evolving XUV radiation in the case of fast (active star), moderate and slow (low activity star) stellar rotation rates. For an Earth-mass planet with an initial atmospheric fraction (ratio between the atmospheric and planetary core masses) of 0.01, the atmosphere was predicted to undergo complete photoevaporation within the initial 300 Myr when the planet orbits a fast-rotating star, whereas it retained nearly its entire atmosphere when orbiting a slowly rotating star. This has important implications for the planetary habitability, as a planet must get rid of its primordial hydrogen atmosphere to become potentially habitable \citep{2009A&ARv..17..181L,2014prpl.conf..883G,2016MNRAS.459.4088O}.

Besides the XUV radiation, stellar winds are as well a byproduct of the magnetic activity of the star, accelerated from the corona by thermal \citep{1958ApJ...128..664P} and MHD wave pressure gradients \citep{2009LRSP....6....3C}, and magnetocentrifugal forces \citep{1976ApJ...210..498B,1967ApJ...148..217W,2017A&A...598A..24J}. Stellar wind properties evolve with rotation and magnetic activity, with denser, faster and hotter winds for magnetically active, fast-rotating stars \citep{2015A&A...577A..27J,2017A&A...598A..24J}.

Stellar winds have the potential to shape the large-scale morphology of the escaping atmospheres, leading to the formation of comet-like tails behind the planet, and confinement of the atmospheres in the substellar line \citep{2015A&A...578A...6M,2019ApJ...873...89M, 2021MNRAS.500.3382C}, which is in agreement with the detected asymmetries in the absorption of the Lyman-$\alpha$ line during planet transit. Stellar winds are able to erode the planet atmosphere, increasing the mass-loss rate \citep{2021ApJ...913..130H,2022MNRAS.510.3039K}, or even reduce it, as shown in the case of stellar wind interacting inside the sonic surface of the planet \citep{2020MNRAS.494.2417V,2021MNRAS.500.3382C}, for either extremely strong stellar winds or faint escaping atmospheres. Furthermore, the interaction of the stellar wind magnetized plasma with the planetary obstacle leads to the formation of potentially detectable characteristic MHD structures, such as bow shock formation in the case of super-fast magnetosonic stellar winds (i.e., when the wind flow velocity exceeds the velocity of the fast magnetosonic MHD wave), or extended magnetospheres around magnetized planets, protecting the planet from additional stellar wind erosion \citep{2015A&A...578A...6M,2019MNRAS.483.2600D,2021MNRAS.508.6001C,2021ApJ...913..130H}. For planets lacking an intrinsic magnetic field, the magnetized stellar wind directly interacts with the planetary ionosphere/conductive planetary surface, leading to the formation of induced magnetospheres \citep{2004AdSpR..33.1905L,2013JGRA..118..321M}, where the magnetic field is piled-up around the conductive obstacle, leading to an increase of magnetic field strength \citep{2023MNRAS.525..286C,2017MNRAS.470.4330E}. The location of these structures is extremely dependent on both the properties of the wind and presence of an atmosphere \citep{2021MNRAS.502.6170C,2023MNRAS.525..286C}. \citet{2021MNRAS.502.6170C} studied the survival of tenuous Earth-like atmospheres against the stellar winds from young to solar-age winds, finding that the bow shock location is sensitive to the properties of the atmosphere, with more massive atmospheres pushing outwards the shock, and stronger winds compressing the shock closer to the planetary surface.  \citet{2017MNRAS.470.4330E} predicted that the magnetic field intensity is enhanced a factor of 2 the local interplanetary magnetic field simulating the interaction of the unmagnetized hot Jupiter HD 209458 b with the stellar wind of its host star. In \citet{2023MNRAS.525..286C}, the interaction of photoevaporated unmagnetized planets with magnetized stellar winds was addressed, finding that the location of the bow shock was sensitive to the magnetic pressure inside the induced magnetosphere, leading to bow shock expansion to larger distances from the planet for low Alfv\'en Mach numbers. 

The interaction between escaping atmospheres and the stellar wind has begun to be addressed in recent years, with a particular emphasis on interactions involving massive escaping atmospheres of hot Jupiters and warm Neptunes \citep{2018MNRAS.479.3115V,2021MNRAS.501.4383V,2021MNRAS.500.3382C,2021MNRAS.508.6001C,2015A&A...578A...6M,2019ApJ...873...89M,2019MNRAS.483.2600D,2022MNRAS.510.3039K}. The relevance of stellar activity in the planet-stellar wind interaction has been addressed in \citet{2022MNRAS.510.3039K}, where the impact of stellar winds from stars of very different ages (and activity) of 50 Myr and 9 Gyr on sub-Neptune planets with similar mass and orbital distances was evaluated. In the case of weak winds from an older star, the interaction between the escaping atmosphere and the stellar wind occurs at greater distances from the planet, resulting in a limited impact on the atmospheric evolution. However, for stronger winds from a younger star, the interaction between the two plasmas occurs closer to the planet, leading to an increased atmospheric mass-loss.

Primordial atmospheres around Earth-mass planets orbiting at further distances are less dense and less expanded in comparison to more massive gaseous exoplanets \citep[e.g. ][]{2013AsBio..13.1011E,2014MNRAS.439.3225L}. The combined evolution of stellar winds and XUV radiation-dependent expanded atmospheres around Earth-mass planets has not been addressed yet, particularly for unmagetized planets, where the escaping atmosphere is not protected by any planetary magnetic field.

In this work, we address this interaction using numerical 3D MHD simulations of the evolution of a photoevaporating hydrogen atmosphere around an Earth-mass planet orbiting at 1.0 au in two different scenarios according to the magnetic activity of the star, considering a fast a and a slow-rotating star, in the time interval between 50 and 500 Myr. The XUV-driven evolution of the escaping H-rich atmosphere was parameterized following the derived mass-loss rates in \citetalias{2015ApJ...815L..12J} in the cited time span, for an Earth-like planet with an initial atmosphere of 0.01 $M_{\oplus}$ (named case C in \citetalias{2015ApJ...815L..12J}). We started our simulations at 50 Myr, where significant atmospheric loss is predicted from the simulations in \citetalias{2015ApJ...815L..12J} in the case of an Earth-like planet orbiting a fast-rotating stars at 1 AU, i.e. when the effects of the XUV-driven photoevaportaion become appreciable. Moreover, at this stellar age, we can find a maximum in rotation and XUV emission of the star, following the evolutionary tracks for solar-like stars by \citet{2015A&A...577L...3T}, used in the 1D atmospheric models in \citetalias{2015ApJ...815L..12J}, both for fast and slow rotators. From this age, a generalized decrease is found in magnitudes defining the characteristic variables of the stellar wind (rotation, magnetic field, mass-loss rate and temperature). Thus, this age range allows us to compare different evolution states according to different stellar activity levels, starting from a maximum value at 50 Myr. We stopped our simulations at 500 Myr, where no atmosphere is found around an Earth-like planet, as predicted in \citetalias{2015ApJ...815L..12J} in the case of a fast-rotating star. We considered same age range for planet orbiting a slow-rotating star for comparison.

The paper is structured as follows: section \ref{sec:paperIII_winds} presents the age-dependent solution of the stellar wind at 1.0 au, employing a 1D Weber \& Davis stellar wind model for both slow and fast-rotating regimes. Section \ref{sec:num_setup_paperIII} outlines our MHD model for simulating the interaction between the stellar wind and the planet, using the MHD module of the \textsc{pluto} numerical code. Section \ref{sec:results_paperIII} provides an overview of the results. Lastly, in Section \ref{sec:discussion_paperIII}, we present our discussion and conclusions.  

\section{Evolution of solar-like stellar winds}
\label{sec:paperIII_winds} 

\begin{table*}
\begin{tabular}{cccccccccccc}
\hline
\hline
\multicolumn{12}{c}{Stellar wind parameters: Fast rotator}                                  \\ \hline\hline
model & Age (Myr) & $v_r $ (km/s) & $v_{\phi}$ (km/s) & n$_p$ (cm$^{-3}$) & $B_r $ (mG) & $B_{\phi}$ (mG) & T (MK) & $M_A$ & $M_{f}$  & $p_{ram}$ (dyn cm$^{-2}$) & SAT \\
\hline

F1    & 50        & 2629       & 67              & 66         & 0.418       & 6.91            & 4.1    & 1.4  & 1.4 & 7.6$\times 10^{-6}$ &yes \\
F2    & 150       & 1948       & 35              & 51         & 0.418       & 3.33             & 4.1    & 1.8  &  1.8  & 3.2$\times 10^{-6}$& yes \\
F3    & 300       & 1525       & 10               & 38         & 0.331       & 0.94             &  3.7    & 4.2  &  3.5  &1.5$\times 10^{-6}$ &no  \\
F4    & 500       & 1175       & 3        & 24         & 0.121      & 0.26             & 2.2    & 9.0  &   5.1 & 5.5$\times 10^{-7}$ & no  \\ 
\hline
\hline
\multicolumn{12}{c}{Stellar wind parameters: Slow rotator}                                  \\ \hline \hline
model & Age (Myr) & $v_r $ (km/s) & $v_{\phi}$ (km/s) & n$_p$ (cm$^{-3}$) & $B_r $ (mG) & $B_{\phi}$ (mG) & T (MK) & $M_A$ & $M_f$ & $p_{ram}$ (dyn cm$^{-2}$) &SAT \\
\hline
S1    & 50        & 1005     & 1.7                & 20          & 0.075       & 0.134             & 1.7    & 13.4  & 5.5 & 3.4$\times 10^{-7}$& no  \\
S2    & 150       & 969      & 1.5                & 17          & 0.065       & 0.107             & 1.6    & 14.3  & 5.5 & 2.6$\times 10^{-7}$ &  no  \\
S3    & 300       & 929      & 1.4                & 31          & 0.058       & 0.092             & 1.5    & 15.3  & 5.6 & 2.2$\times 10^{-7}$&no  \\
S4    & 500       & 804      & 1.4                & 29          & 0.054       & 0.085            & 1.1    & 13.9  & 5.5 & 1.6$\times 10^{-7}$& no \\
\hline
\hline
\end{tabular}
\caption{Stellar wind parameters at 1.0 au for each considered stellar age, both for rapidly rotating (top panel) and slowly rotating stars (bottom panel). Values for radial ($v_r$) and azimuthal ($v_{\phi}$) components of the velocity field, proton density number $n_p$,  radial ($B_r$) and azimuthal ($B_{\phi}$) components of the magnetic field, temperature $T$, the Alfv\'en and fast magnetosonic Mach numbers, $M_A$ and $M_f$ respectively, and the characteristic dynamic (ram) pressure of the stellar wind are given according to the WD stellar wind model solution. The last column indicates whether the star is in either the saturated regime or the unsaturated regime.}
\label{tab:sw_param}
\end{table*}

Main ingredients driving stellar winds, named stellar rotation and magnetic activity evolve throughout the stellar lifetime \citep{2021LRSP...18....3V,2021arXiv210511243J}, resulting in the subsequent evolution of stellar winds.

Here, we obtained the activity-dependent stellar wind parameters at 1.0 au for a solar-like star ($1 M_{\odot}, 1R_{\odot}$) using the 1D axisymmetric Weber \& Davis \citep[][hereafter WD]{1967ApJ...148..217W} solar wind model, scaled to different stellar activity levels, distinguishing between fast-rotating (active) and slow-rotating (low-activity) stars, at different stellar ages of 50, 150, 300 and 500 Myr.\\
The WD model provides a simplified approach to predict stellar wind parameters at different distances from the stellar corona while accounting for magnetocentrifugal effects in the acceleration of stellar winds, which are particularly relevant, if not the main wind acceleration mechanism for fast-rotating stars. The WD model assumes a split monopole geometry as an approximation for the coronal magnetic field topology of the star. In this sense, the complex 3D coronal magnetic field of the star is not taken into account, leading to a simplified stellar wind distribution, where the three critical surfaces of the wind, defined as the critical points where the stellar velocity flow equals the velocity of the slow magnetosonic, Alfv\'en and fast magnetosonic magnetohydrodynamic waves, appear as perfect circumferences in the stellar disk.

The topology of the Alfvén surface is critical in the stellar wind-planet interaction, as very different interactions take place under the sub-alfvénic and the super-alfvénic regimes \citep[see e.g.][]{strugarek2021physics}. However, in our study, the stellar wind WD solution for the coronal parameters considered in the case of fast- and slow-rotating stars of different ages, predicts the Alfvén surface well below the fixed orbital distance of 1 AU. Consequently, no changes in the wind regime are expected in the considered scenario, where the planet will orbit outside the Alfvén surface along its entire orbital path.

Even though the WD 1D approximation is far from the sophisticated 3D simulations of the corona of low-mass stars \citep[e.g.][]{2023A&A...678A.152V,2022MNRAS.512.4556S,2016ApJ...833L...4G}, it can be very useful to perform a parametric study considering different coronal conditions, as well as other more simplified stellar wind 1D models \citep[e.g.][]{2022MNRAS.510.3039K,2021MNRAS.500.3382C}, often used to simulate the stellar wind-planet interaction.

The WD model requires six input parameters to be defined at the base of the heated corona, namely the stellar radius, the stellar mass, the rotation period of the star, the mass-loss rate, the radial component of magnetic field at the base of the corona and the wind base temperature. The MHD equations describing a fluid accelerated due to thermal pressure gradients and magnetocentrifugal forces are solved numerically using the approach described in \citet{2005A&A...434.1191P}, predicting an expanded coronal wind under the isothermal approximation for a fully ionized plasma. 

Here, we consider the rotation rates from the fast and slow-rotating evolutional tracks derived by \citet{2015A&A...577L...3T}. Based on observations of $\sim$2000 stars within clusters with ages spanning from 20 to 500 Myr, \citet{2015A&A...577L...3T} obtained the evolutionary trends of rotation, and X-ray and EUV luminosities. Three different rotation trends are derived according to the considered initial rotation rate of the star, $\Omega_0$, distinguishing between fast ($\Omega_0 = 45.6 \Omega_{\odot}$), moderate ($\Omega_0 = 6.2 \Omega_{\odot}$) and slow ($\Omega_0 = 1.8 \Omega_{\odot}$) rotator branches, where $\Omega_{\odot} = 2.9 \times 10^{-6}$ rad s$^{-1}$ is the current rotation rate of the Sun. In the fast-rotating regime, rotation rates range between 0.3 and 4.3 days in the 50-500 Myr age range, while rotation rates are between 6 and 8.6 days in the slow-rotating regime in the same period. 

At high-spin rates, X-ray and EUV radiation of the star saturate (i.e. become independent of rotation). According to their model, the saturation regime ends at $\sim6$ Myr for slow rotators when the rotation rate of the star is lower than the saturation threshold $\Omega_{sat} \sim 2.0\Omega_{\odot}$, and at $\sim 220$ Myr for fast rotators, where the stellar rotation rate becomes lower than $\Omega_{sat} = 15 \Omega_{\odot}$. Therefore, in our considered stellar age range, a slow-rotating star will remain in the unsaturated regime, while for a fast-rotating star, saturation regime has to be considered at 50 and 150 Myr.

For the mass-loss rate of the star we follow the relation given in \citet{2015A&A...577A..27J} \citep[also used in the rotational models of][]{2015A&A...577L...3T}):
\begin{equation}
    \dot{M_{\star}} \propto R_{\star}^2 \Omega^{1.33} M_{\star}^{-3.36}
\end{equation}
valid for the unsaturated regime using a reference value for the solar mass-loss rate of $1.4\times10^{-14}M_{\odot}/yr$. This relation predicts an increased mass-loss rate for active, fast-rotating unsaturated stars. For saturated stars, we replaced the angular velocity $\Omega$ of the star by the saturated value of 15$\Omega_{\odot}$ \citep{2015A&A...577A..28J}. \citet{2017A&A...598A..24J} studied the behavior of mass-loss rate for fast-rotating stars in the saturated regime conducting a stellar wind 1D MHD model, using two different approaches for the base wind temperature $T_0$ (named model A and model B in \citet{2017A&A...598A..24J}). In model A, it is assumed that the coronal temperature scales with the X-ray flux of the star, that is at the same time dependent on the rotation rate of the star. This model predicts hotter winds for more rapidly rotating stars. In model B, the sound speed at the base of the wind is assumed to be a constant fraction of the escape velocity. Under this approximation, the wind temperature is independent on the rotation rate of the star. Following model A, that is the preferred method in \citet{2017A&A...598A..24J}, they found that the mass-loss rate is kept constant from $\Omega_{sat}$, and rapidly increases from $\Omega=100\Omega_{\odot}$ (see their Figure 2) due to the magnetocentrifugal effects. In concordance, in this work we use a value of $\dot{M_{\star}}=1\times10^{-12}$ $M_{\odot}/yr$ when stellar angular velocities $\gtrsim 100 \Omega_{\odot}$, a value reached for a fast-rotating stars at 50 Myr \citep{2015A&A...577L...3T}. For a fast-rotating star, the derived mass-loss rates at each considered stellar age between 50 and 500 Myr range between 1$\times10^{-12}$ and 2$\times10^{-13}$ $M_{\odot}$/yr, and between 1.4$\times10^{-13}$ and 8.7$\times10^{-14}$ $M_{\odot}$/yr in the slow-rotating star case. 
 
Coronal temperatures are derived for each considered stellar age following the universal relation of \citet{2015A&A...578A.129J} for low-mass stars, showing a clear dependence with the X-ray flux of the star as:
\begin{equation}
    T_{cor} = 0.11F_X^{0.26}
\end{equation}

Here, X-ray fluxes are from \citet{2015A&A...577L...3T}. The base-wind temperature of the stellar wind is known to correlate with the coronal temperature. We scaled down the coronal temperature by a factor of 1.36 \citep{2018MNRAS.476.2465O}. Base wind temperatures in the fast-rotating regime range between 7.5 and 4 MK in the 50-500 Myr age range, and between 3.1 and 2.5 MK in the slow-rotating regime. The isothermal assumption used in the WD model results in unrealistic high temperature values at distant points from the star's surface. To address this, we adopted a polytropic for the calculation of the wind temperature at each orbital distance by employing a constant polytropic index of $\gamma$ = 1.1 \citep[as justified in ][]{2021MNRAS.502.6170C}. 

The equatorial magnetic field of the star is scaled from the empirical trends of \citet{2014MNRAS.441.2361V}:
\begin{equation}
   B_{\star} \propto R_o^{-1.38}
\end{equation}
where $R_o$ is the Rossby number of the star at a given age, defined as $R_o = P_{rot}/\tau_c$, where $P_{rot}$ is the rotation period of the star and $\tau_c$ is the convective turnover time. In this work, we used the $\tau_c$-stellar age relations given in \citet{2010A&A...510A..46L} for a solar-mass star. Magnetic field of the star saturates at a Rossby number of 0.13 \citep{2018MNRAS.479.2351W,2011ApJ...743...48W}. For a fast-rotating star in the 50-500 Myr age range, the magnetic field ranges between 17 and 7.3 G, while for a slow-rotating star, the radial magnetic field at the surface of the star is decreased with values between 3.1 and 2.5 G. 

WD stellar wind parameters at 1.0 au in the case of a slow and a fast rotators are given in Table \ref{tab:sw_param} for each considered stellar age. The WD model predicts faster, hotter, more magnetized and denser winds for fast rotators, in comparison to the slow rotator case. The main parameters of the wind decrease with the stellar age as expected in both analyzed cases. Our stellar wind parameters are consistent with those obtained in the 1D MHD model of \citet{2017A&A...598A..24J}, where the effect of the magnetocentrifugal forces was evaluated. As a reference, they obtained a wind accelerated up to 2600 km s$^{-1}$ with a proton density of $\sim 60$ cm$^{-3}$ for a star rotating at 100 $\Omega_{\odot}$, values that are in agreement with our values for F1 model, were the rotation rate of the star is $\sim100 \Omega_{\odot}$ (see Table \ref{tab:sw_param}).

For each evolutionary stage, winds are super-alfv\'enic, super-fast magnetosonic ($M_A,M_{fm}>1.0$, where $M_A$ is the Alfv\'en Mach number defined as the ratio between the flow velocity and the Alfv\'en velocity $V_A = B/\sqrt{4\pi\rho}$, and $M_{fm}$ is the fast magnetosonic Mach number, defined as the ratio between the flow velocity and the velocity of the fast magnetosonic MHD wave, defined as $v_{fm} = \sqrt{c_s^2 + v_A^2}$, where $c_s$ is the sound speed) at 1.0 au, both for fast and slow rotators. According to this characteristic, a bow shock is expected to form in front of the planetary obstacle, whose location is expected to be at further distances for lower alfv\'enic Mach numbers \citep{2023MNRAS.525..286C}.

\section{Numerical setup}
\label{sec:num_setup_paperIII}
The interaction of the stellar wind and the escaping primordial atmosphere of an Earth-like planet is studied through numerical, ideal MHD 3D simulations. We conducted our simulations in 3D spherical coordinates using the MHD module of the single-fluid \textsc{pluto} code \citep{2007ApJS..170..228M}, where the spatial and temporal evolution of the MHD equations of continuity, momentum, energy, and magnetic induction is performed: 
\begin{equation}
    \frac{\partial \rho}{\partial t}+{\nabla} \cdot (\rho \textbf{v})=0 
    \label{masa}
\end{equation}

\begin{equation}
    \frac{\partial(\rho\mathbf{v})}{\partial t} + \nabla\cdot\left[\rho\mathbf{v}\mathbf{v}-\frac{\mathbf{B}\mathbf{B}}{4\pi}+p_T\mathbf{I}\right]^T= \rho(\mathbf{g}+\mathbf{F}_{Cor}+\mathbf{F}_{cen})
\end{equation}

\begin{equation}
    \frac{\partial E_T}{\partial t} + \nabla \cdot \left[(E_T+p_T)\mathbf{v}-\frac{\mathbf{B}}{4\pi}(\mathbf{v}\cdot\mathbf{B})\right] = \rho\left( \mathbf{g}+\mathbf{F}_{cen}\right)\cdot\mathbf{v}
\end{equation}
    
\begin{equation}
    \frac{\partial \mathbf{B}}{\partial t} + \nabla \times \left(\mathbf{B}\times\mathbf{v}\right) = 0
\end{equation}

The mass density, velocity, magnetic field and gravitational acceleration (including both the acceleration due to the planetary and stellar gravity) vectors are denoted as $\rho$, $\mathbf{v}, \mathbf{B}$ and $\mathbf{g}$, respectively. The pressure term $p_T=p+\frac{\mathbf{B}^2}{8\pi}$ includes the corresponding thermal ($p$) and magnetic pressure contributions. The total energy of the system, $E_T$, is defined as $E_T=\rho \epsilon + \frac{\rho \mathbf{v}^2}{2}+\frac{\mathbf{B}^2}{8 \pi}$, where $\epsilon$ is the internal energy per mass.\\

An ideal equation of state is solved in conjunction with the ideal MHD equations, using an quasi-isothermal approach with a polytropic index $\gamma$ = 1.01:
\begin{equation}
    \rho\epsilon = \frac{p}{\gamma-1}
\end{equation}
The selected $\gamma\sim1$ polytropic index ensures an almost constant temperature for the planetary wind, as it will be described below.

To take into account the effects of the orbital motion of the planet around the star, we solved the ideal MHD equations in a frame of reference co-rotating with the planet, considering a constant angular velocity $\Omega=\sqrt{GM_{\star}/d^3}$, where $d$ the orbital radius of the planet. We modified the right-hand side of the momentum and energy equations, incorporating the corresponding inertial terms of the Coriolis and centrifugal accelerations:
\begin{equation}
    \mathbf{F}_{Cor} = -2(\mathbf{\Omega}\times \mathbf{\vec{v}}) 
\end{equation}
\begin{equation}
    \mathbf{F}_{cen} = -\mathbf{\Omega}\times(\mathbf{\Omega}\times \mathbf{\vec{v}}) 
\end{equation}

However, for the distance of 1.0 au considered in this study, the effects of centrifugal and Coriolis forces are irrelevant in our simulations.

Under the ideal MHD approximation (i.e., the resistivity term is neglected in the induction equation), both atmospheric and stellar winds are described as a fully ionized, perfectly conducting plasma. This means that the magnetic field is not able to penetrate inside regions where the magnetic field magnitude is initially defined to zero. This configuration allows to reproduce the interaction of the magnetized stellar wind with the upper highly ionized atmosphere, as expected due to the high XUV flux at early ages.

The  Harten, Lax, Van Leer (hll) approximate Riemann solver has been selected for numerical flux computation in the interface midpoint. A linear reconstruction associated with a diffusive minmod limiter in all variables is used. A third order Runge-Kutta scheme is implemented in the simulations for time evolution. To ensure the solenoidal condition of the magnetic field, $\nabla \cdot \textbf{B} = 0$, we tested two different methods: (i) the extended generalized Lagrange Multiplier \citep[GLM, ][]{2010JCoPh.229.5896M,2010JCoPh.229.2117M} formulation, and (ii) the Eight waves formalism \citep{1994arsm.rept.....P}. Even though the GLM method has been proven to be more accurate and robust for simulations including low-$\beta$ plasma, this formalism lead to excessive magnetic field turbulence at grid regions where simultaneously the resolution is low and strong shocks are formed. This leads to unstable regions close to the shock formation, altering the steady state position of these shocks. For this reason, we performed our simulations using the eight-wave formalism for the magnetic field diverge control, leading to a more stable interaction in low-resolution regions. Near-zero values for the magnetic field divergence are obtained in the whole computational grid under the cited approximation. However, with exception of the magnetic field turbulence, both methods lead to compatible results, as we tested in preliminary simulations (see Appendix \ref{ap1}).\\

The spherical 3D computational grid is made of 738 points in the radial direction $r$, 96 for the polar angle $\theta$, and 192 for the azimuthal angle $\phi$, with $r$ $\in [2.5, 60R_p]$, $\theta$ $\in[0,\pi]$ and $\phi$ $\in[0,2\pi]$. In the corresponding Cartesian coordinate system, the X-axis direction is pointing from the planet toward the star, the Y-axis is chosen to be in the ecliptic plane and the Z-axis is parallel to the ecliptic pole, adopting the Geocentric Solar Ecliptic System (GSE) of coordinates, where the planet is located at the center of the grid (hereafter, we will refer to X,Y,Z to the GSE system described coordinates). Maximum resolution of $\sim$0.08$R_p$ is found in the radial, azimuthal and polar coordinates close to the planet (inner) boundary. The resolution naturally decreases for increasing radial distance from the center of the grid, with a maximum cell side of 1.9$R_p$ in the polar and azimuthal components.

\begin{figure}
    \centering
    \includegraphics[scale=0.33]{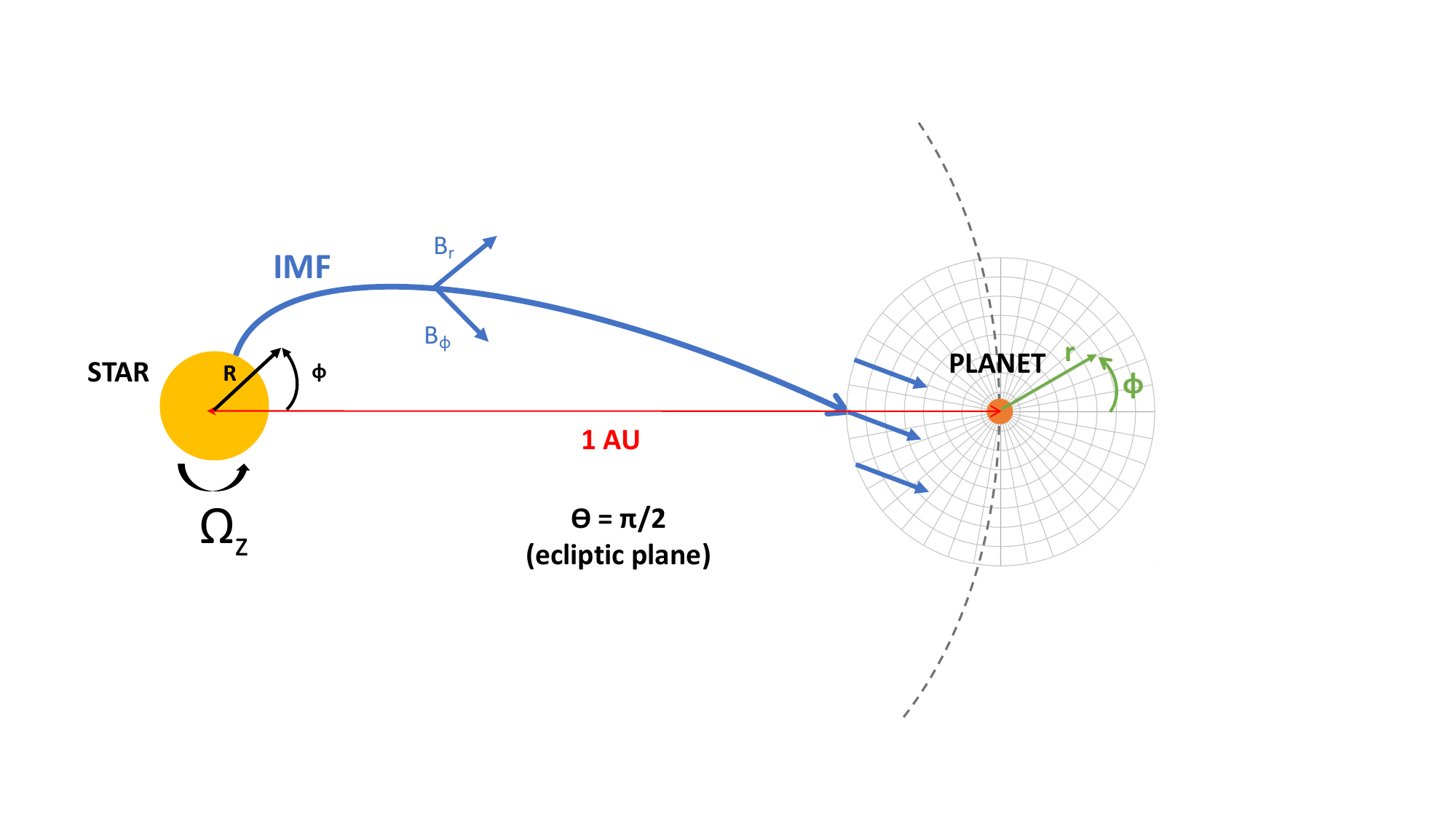}
    \caption{Scheme of the interplanetary magnetic field (IMF) orientation, obtained from the solution of the WD model, as defined at the outer boundary of the spherical computational domain. The spherical coordinates are shown for both systems centered in the host star and in the planet in the ecliptic plane ($\theta=\pi/2$). The dashed black line represents the orbital path of the planet.}
    \label{fig:mf_scheme}
\end{figure}

Fixed parameters of density and temperature are defined at the inner boundary at 2.5 $R_p$ to reproduce the escaping atmosphere of the planet, and are described in detail in the next subsection. The stellar wind is injected from the outer boundary of the domain at $r=60R_p$ for $\pi/2<\phi<3\pi/2$, fixing the stellar wind components of the magnetic field and velocity vectors, gas pressure and density as constant boundary conditions, following the stellar wind solution obtained at each stellar age with the WD model, described in the last section. An scheme representing the orientation of the magnetic field lines as injected in the spherical computational domain is shown in \ref{fig:mf_scheme}.  Outflow boundary conditions are defined at the outer boundary for $-\pi/2<\phi<\pi/2$.

\subsection{Planetary outflow}
XUV stellar radiation heats and photoevaporates planetary hydrogen atmospheres during the early stages of a planet life, leading to a gradual loss of the atmospheric envelope and reducing the atmospheric escape rate. We modeled the evolution of the XUV-heated escaping atmosphere of a terrestrial planet orbiting a solar-like star at 1.0 au based on the atmospheric parameters derived in the 1D hydrodynamic model of \citetalias{2015ApJ...815L..12J}. Using the Versatile Advection Code (VAC, Toth et al. 1996), \citetalias{2015ApJ...815L..12J} performed a 1D hydrodynamic upper atmosphere model including the gravity of the planet and XUV heat deposition to calculate the atmospheric loss as a function of the incident XUV radiation in the case of Earth-mass (between 0.5 and 5.0 $M_{\oplus}$) planets, along with the density, temperature, and velocity structure of the planet outflow. 

In their simulations, they considered a fixed equilibrium temperature of 250 K and number density of 5$\times 10^{12}$ cm$^{-3}$ at the base of the planetary wind, defined at a distance $R_0>R_{core}$ from the center of the planetary core of radius $R_{core}$. The base wind $R_0$ is set to be dependent on the atmospheric mass fraction ($f_{atm} = M_{atm}/M_{p}$) as: 
\begin{equation}\label{eq:z0}
    log\left(\frac{R_0}{R_{core}}\right) = (2.5f_{atm}^{0.4}+0.1)\left(\frac{M_{p}}{M_{\oplus}}\right)^{-0.7}
\end{equation}
with more massive atmospheres leading to larger wind base $R_0$. The atmospheric base decreases as the atmosphere is photoevaporated by the XUV radiation of the star. 

The atmospheric mass-loss rate was adjusted as a function of the incident XUV radiation of the star for a given atmospheric $z_0 = R_0-R_{core}$, as 
\begin{equation}\label{eq:mass_loss}
    \dot{M}_{atm} = 1.858\times10^{31}m_H z_0^{0.464}(log F_{XUV})^{4.093 z_0^{-0.022}}
\end{equation}
predicting higher atmospheric escape for highly XUV-irradiated planets. Using the $F_{XUV}$ evolutional tracks for fast, moderate and slow-rotating solar-like stars derived in \citet{2015A&A...577L...3T}, \citetalias{2015ApJ...815L..12J} used equations \ref{eq:mass_loss} and \ref{eq:z0} to integrate the atmospheric mass for different initial atmospheric mass fractions and planetary masses, predicting a progressive loss of the primordial atmospheres. The temperature profiles of the escaping atmospheres are dependent on the incident XUV flux, finding atmospheres heated up to 1$\times10^4$ K at larger distances ($\sim 20 R_p$) for an incident flux of 5000 erg s$^{-1}$ cm$^{-2}$.

\begin{table*}
\caption{Atmospheric parameters adopted in our simulations. The atmospheric fraction $f_{atm}$, the base wind $R_0$ and the atmospheric mass-loss rate $\dot{M}$ are obtained from 1D models by \citetalias{2015ApJ...815L..12J}. Las two columns specify the planetary wind temperatures and number densities defined at the inner boundary in our simulations.}
\begin{tabular}{cccccccc}
\hline
\hline
\multicolumn{8}{c}{Atmospheric parameters: Fast rotator}                                  \\ \hline\hline
model & Age (Myr) & $F_{XUV}$ (erg s$^{-1}$ cm$^{-2}$) &  $f_{atm} $ & $R_0$ (R$_p$) & $\dot{M}$ ($1\times 10^{9}$ g/s)  & T (K) & $n_0$ (cm$^{-3}$) \\
\hline

F1    & 50        & 3443 & $1\times10^{-2}$       & 2.4         & 6.1 &   5000       & $6.5\times10^{8}$            \\
F2    & 150       & 3443 & $5\times10^{-3}$       & 2.0     & 5.5  & 5000       & $6.0\times10^{8}$             \\
F3    & 300       & 2268 & $1\times10^{-3}$       & 1.7               & 3.9         & 4000       & $4.7\times10^{8}$              \\
F3b    & 300       & 2268 &  $1\times10^{-6}$       & 1.7               & 1.9         & 4000       & $2.0\times10^{8}$              \\
F4    & 500      &  542  &    -   &         -       &     -    & 2000     &  $1.0\times10^{6}$              \\ 
\hline
\hline
\multicolumn{8}{c}{Atmospheric parameters: Slow rotator}                                  \\ \hline \hline
model & Age (Myr) & $F_{XUV}$ & $f_{atm} $ & $R_0$ (R$_p$) & $\dot{M}$ ($1\times 10^{9}$ g/s)  & T (K) & $n_0$ (cm$^{-3}$) \\
\hline
S1    & 50        & 224  & $1\times10^{-2}$   & 2.4       & 1.2         & 1200          & $2.6\times10^{8}$                     \\
S2    & 150     & 182  &    $1\times10^{-2}$   & 2.4      & 1.0          & 1100          & $2.5\times10^{8}$                     \\
S3    & 300     & 132 &    $8\times10^{-3}$   & 2.3        & 0.8        & 1000          & $2.1\times10^{8}$                       \\
S4    & 500     & 95   &    $7\times10^{-3}$   & 2.2      &    0.5      & 900          & $1.5\times10^{8}$                  \\
\hline
\hline
\end{tabular}
\label{tab:atm_param}
\end{table*}

In the present study, we considered the time-dependent integrated atmospheric mass fractions of \citetalias{2015ApJ...815L..12J} in the case of an Earth-mass planet orbiting at 1.0 au, considering an initial atmospheric mass fraction of 0.01 (named case C in \citetalias{2015ApJ...815L..12J}, see their Figure 5). This initial atmospheric fraction is consistent with the predicted atmospheric mass accreted from the protoplanetary disk for Earth-mass planets \citep{2020SSRv..216..129O}. Moreover, \citetalias{2015ApJ...815L..12J} predicted drastically different atmospheric evolution scenarios for that initial fraction, considering the effects of fast and slow-rotating stars. Using equations [\ref{eq:mass_loss}] and [\ref{eq:z0}] and the $F_{XUV}$ evolutional trends of \citet{2015A&A...577L...3T}, we calculated the corresponding mass-loss rates and atmospheric base heights for each test case. The values of $f_{atm}$, $z_0$ and the derived mass-loss rate $\dot{M}$ are given in Table \ref{tab:atm_param}, for both fast and slow rotator regimes. Due to the decrease in stellar $F_{XUV}$ with stellar age, the atmospheric mass-loss rate is decreased, and the atmospheric mass fraction is as well reduced with time due to atmospheric erosion.  

As the complete photoevaporation is predicted from the evolution models of \citetalias{2015ApJ...815L..12J} around $\sim$300 Myr for a planet orbiting a fast-rotating star, we included a test case at 300 Myr (named F3b), where the stellar wind and radiation parameters are fixed, but the atmospheric mass fraction (and consequently the mass-loss rate) is considerably reduced up to three orders of magnitude in comparison to F3 model, consistent with massive atmospheric losses predicted in \citetalias{2015ApJ...815L..12J}. In model F4, corresponding to an age of 500 Myr, the atmospheric fraction is reduced to $\sim$zero, reproducing a non-atmosphere planet. For a planet orbiting a slowly rotating star, the evolution of atmospheric mass-loss and the remaining fraction of atmosphere remains relatively constant, without undergoing significant changes as observed in the case of a fast-rotating star.

To model a heated planetary outflow characterized by a specific mass-loss rate in our MHD code, we modeled the planetary wind as a Parker-like isothermal flow, considering the gas to be composed entirely by Hydrogen, where the planetary wind is only accelerated by the contribution of thermal pressure gradient, and balanced with the gravity of the planet \citep[see e.g.][]{2022A&A...659A..62D,2021ApJ...913..130H,2021MNRAS.508.6001C}. At the inner boundary of our spherical domain, fixed in all models at 2.5 $R_p$, we defined fixed values for density and temperature to match the mass-loss rates and velocity profiles and temperatures obtained in the 1D HD numerical code of \citetalias{2015ApJ...815L..12J} at each considered stellar age, both for fast and slow rotator regimens. The velocity in the radial direction at the inner boundary is set to be reflective, having the same value but opposite sign compared to the adjacent cells, resulting in near zero velocities at the inner boundary. The azimuthal and polar component of the velocity field are set to zero. Same configuration is defined for the magnetic field radial, azimuthal and polar components. 

As constant values of temperature and density are fixed at the whole internal boundary (i.e. the wind base), the planetary wind has spherical symmetry. Most works dealing with the planet-stellar wind interaction adopt this kind of approximation \citep[see e.g.][]{2022MNRAS.509.5858H, 2022ApJ...934..189C, 2021MNRAS.500.3382C}. These works and the one presented here assume several simplifications that may affect the steady state of the planetary outflow when interacting with an incident stellar wind, as a quasi-isothermal assumption \citep[neglecting the nighside of the planet, that can affect the mass loss rate XUV driven espace of the planetary atmosphere, as described in][]{2017MNRAS.466.2458C}, charge-exchange reactions, and the effects of radiation pressure, that are found significantly contribute to the acceleration of neutral hydrogen in the planetary atmosphere \citep{2019MNRAS.487.5788E}. The incorporation of these effects will be analyzed in future work.

Values of density at the inner boundary of our computational domain leading to the corresponding XUV-driven mass-loss rates from \citetalias{2015ApJ...815L..12J} are given in Table \ref{tab:atm_param}. Density values at the fixed $R_0=2.5$ R$_p$ boundary decrease with stellar age and atmospheric mass fraction, reproducing an atmosphere that becomes less dense due to thermal losses. The selected temperatures at the base of the wind are consistent with the predicted flow temperatures of the HD model of \citetalias{2015ApJ...815L..12J} at large distances. In models F1 and F2, corresponding to a fast-rotating star in the saturated regime, the temperature is fixed at 5000 K. The temperature is slightly decreased to 4000 K in models F3 and F3b according to the reduced XUV radiation, and dropped to 1600 K in model F4. In the slow-rotating regime, temperatures decrease with age ranging from 1200 K to 900 K, corresponding to a lower XUV emission of the star in comparison to the fast-rotating regime. In addition, the quasi-isothermal approximation assumed in our MHD simulations guarantees that the defined temperature at the base of the planetary wind is preserved at larger distances from the planet. 
 
It's important to remark that in the HD model of \citetalias{2015ApJ...815L..12J}, the atmospheric base $R_0$ is defined as the distance from the center of the planet where density is $5\times10^{12}$ cm$^{-3}$. Using this density value at the inner boundary in our model results in excessive mass-loss rates of several orders of magnitude higher than those predicted by \citetalias{2015ApJ...815L..12J}. The fixed value of $R_0=2.5$ R$_p$ used in our model is located slightly further away than any of the $R_0$ values predicted by equation [\ref{eq:z0}] (see Table \ref{tab:atm_param}). Atmospheric density profiles obtained in \citetalias{2015ApJ...815L..12J} show a rapid decrease within the first planetary radii, diminishing by several orders of magnitude (see their Figure 1). The chosen density values at the base of our model range between $6.5-1.5\times10^8$ cm$^{-3}$, which are consistent with the atmospheric density values slightly above the defined wind base obtained in \citetalias{2015ApJ...815L..12J}.

As initial conditions at t$=0$, we initialize the velocity field as a Parker-like isothermal profile, $v(r) = a(1 - R_0/r)^b$, where $a$ and $b$ parameters are adjusted in each case to provide a fast convergence into the steady-sate solution of the planetary wind. Density profile in the initial conditions is defined to preserve mass conservation, $n(r) = n_0(R_0/r)^2(v_0/v(r))$. 

Initially, we run the simulations in absence of the stellar wind, leading the planetary wind converge to the steady state. Once the planetary wind reached stable density and velocity profiles, we launch the stellar wind from the outer boundary of the domain.

\section{Results}
\label{sec:results_paperIII}
\subsection{Planetary winds}
\begin{figure*}
    \centering
    \includegraphics[scale=0.57]{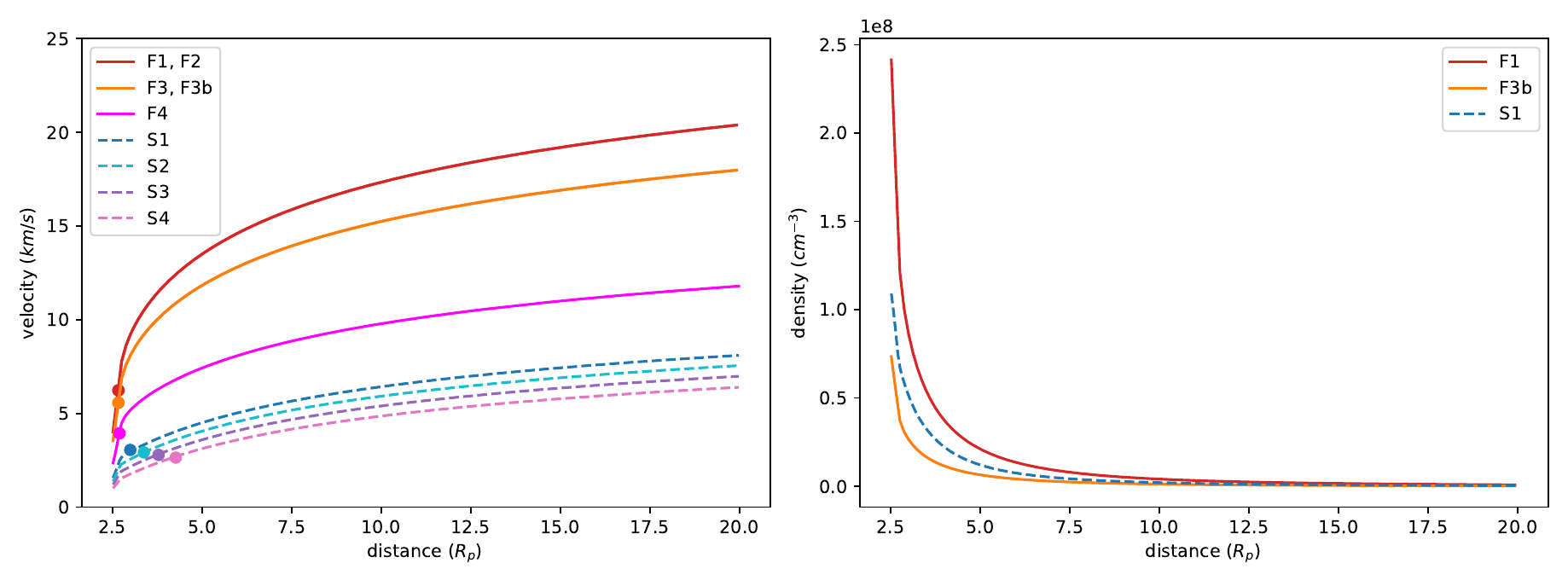}
    \caption{Left panel: Velocity profiles of the planetary outflow in absence of the stellar wind, obtained in the steady state solution of our MHD simulations. Planets orbiting a fast (slow) rotating star are represented by solid (dashed) lines. The solid dot represents the location of the sonic surface for each model. Right: Density profiles of the planetary outflow for models F1 (fast-rotating, dense atmosphere), F3b (fast-rotating, faint atmosphere) and S1 (slow-rotating, dense atmosphere).}
    \label{fig:planetary_winds}
\end{figure*}
Planetary wind velocity and density profiles in absence of the stellar wind at the simulation steady-state are given in Fig. \ref{fig:planetary_winds}. The flow is ejected radially from the inner boundary, keeping this distribution at larger distances. Higher velocities are found for planets orbiting fast-rotating stars as higher temperatures are considered at the planetary wind base, corresponding to higher stellar XUV fluxes. The planetary wind resulting from our simulations compare well with \citetalias{2015ApJ...815L..12J} 1D results. For a planet orbiting a fast-rotating star in the saturated regime (models F1 and F2 at 50 and 150 Myr respectively), planetary winds reach 20 km/s at 20 R$_p$. The wind velocity is slightly lower in models F3 and F3b (unsaturated fast-rotating star of 300 Myr). The planetary wind velocity is dropped to 10 km/s at 20 R$_p$ in model F4, as lower values of XUV radiation are considered. Finally, for planets orbiting a slow-rotating star (dashed lines in Fig. \ref{fig:planetary_winds}), wind velocities at 20 R$_p$ are between 8 and 6 km/s. 

As a consequence of XUV-driven atmospheric mass-loss, density profiles are decreased with stellar age (see models F1 ($f_{atm}$ = 0.01) and F3b ($f_{atm}=1\times10^{-6}$) for comparison). Even though models F1 and S1 have similar initial atmospheric mass fractions, the planetary atmosphere in model F1 is more inflated in comparison to model S1, resulting in higher densities at far distances from the planetary base, as shown in the right panel in Fig. \ref{fig:planetary_winds}. 

Planetary winds in all considered configurations start from subsonic velocities, reaching supersonic velocities at close distances from the planet from the sonic surface (represented by a solid dot in left panel in Fig. \ref{fig:planetary_winds}). The location of the sonic spherical surface is found at 2.7 $R_p$ in the fast-rotating regime (F$_{XUV}$ between 3443 and 2268 erg s$^{-1}$ cm$^{-2}$) for ages between 50 and 300 Myr, while this critical surface is found slightly further away at 2.8 $R_p$ in the F4 model (F$_{XUV}$=542 erg s$^{-1}$ cm$^{-2}$). The sonic surface is found at larger distances from the planet in the slow-rotating regime (F$_{XUV}$ between 224 and 95 erg s$^{-1}$ cm$^{-2}$), ranging from 3.0 R$_p$ up to 4.7 R$_p$ with increasing stellar age. Sonic surface locations differ from those predicted in \citetalias{2015ApJ...815L..12J}, found at slightly larger distances from the planet in comparison to our MHD planetary wind models (in \citetalias{2015ApJ...815L..12J}, the sonic surface is found approximately at 3.0R$_p$, 4.5R$_p$ and 9.5R$_p$ from the base of the wind, for atmospheres irradiated with fluxes of 5000, 500 and 50 erg s$^{-1}$ cm$^{-2}$, respectively) due to the isothermal approximation assumed in our numerical simulations, in conjunction with a high temperature defined at inner boundary of our simulation domain. However, as we show later, the interaction with the stellar winds takes place at larger distances from the sonic point location predicted by \citetalias{2015ApJ...815L..12J}, not affecting the planetary mass-loss rate from the wind base.

\subsection{Stellar wind - escaping atmosphere interaction outline}
\begin{figure*}
    \centering
    \includegraphics[scale=0.48]{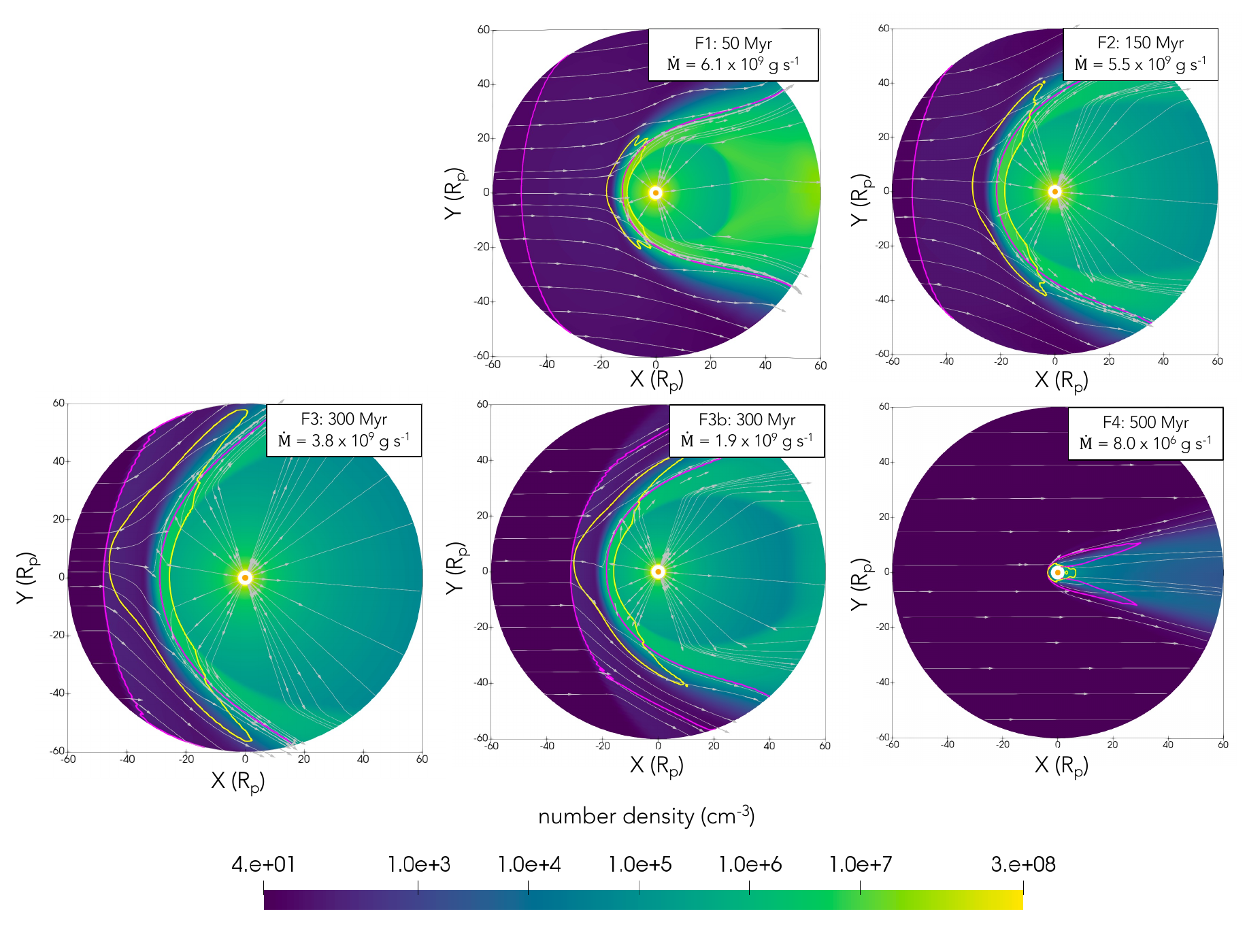}
    \caption{Density distribution map in the ecliptic (XY) plane for an Earth-like planet orbiting a fast-rotating star in the age range from 50 to 500 Myr. Grey thin lines represent the flow stream lines. Pink and yellow thick lines are for alfv\'enic and sonic surfaces, respectively.}
    \label{fig:fast_rotator_density}
\end{figure*}

\begin{figure*}
    \centering
    \includegraphics[scale=0.54]{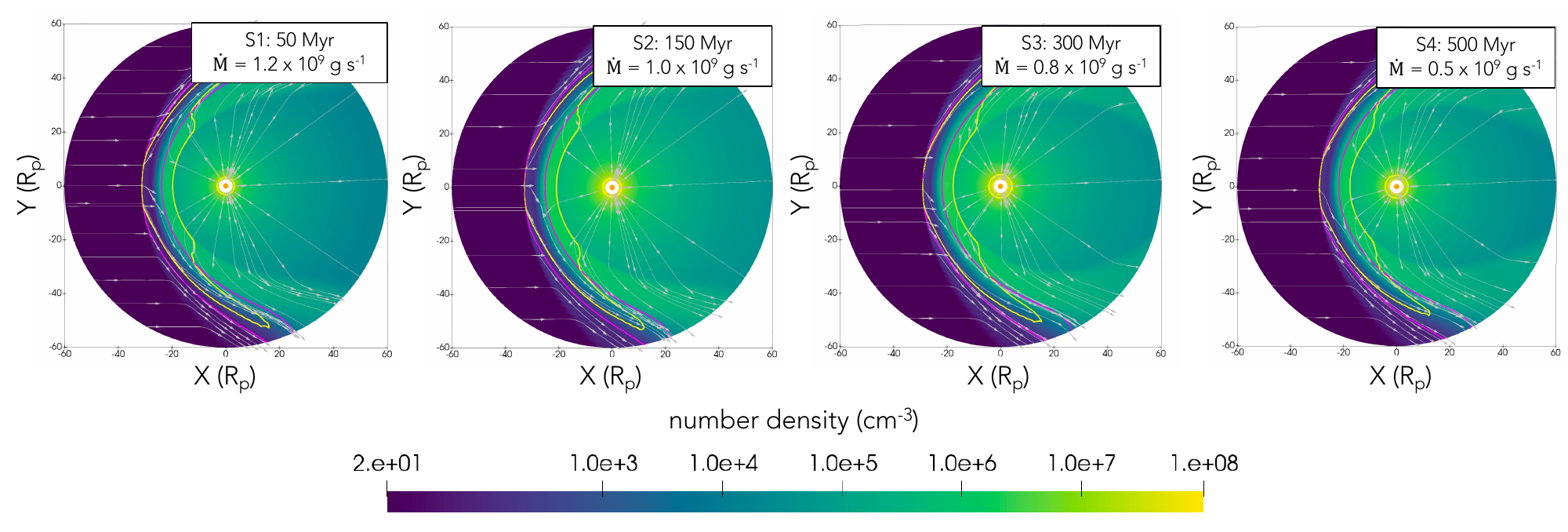}
    \caption{Same as Fig. \ref{fig:fast_rotator_density}, but considering models in the stellar slow-rotating regime. }
    \label{fig:slow_rotator_density}   
\end{figure*}

\begin{figure*}
    \centering
    \includegraphics[scale=0.6]{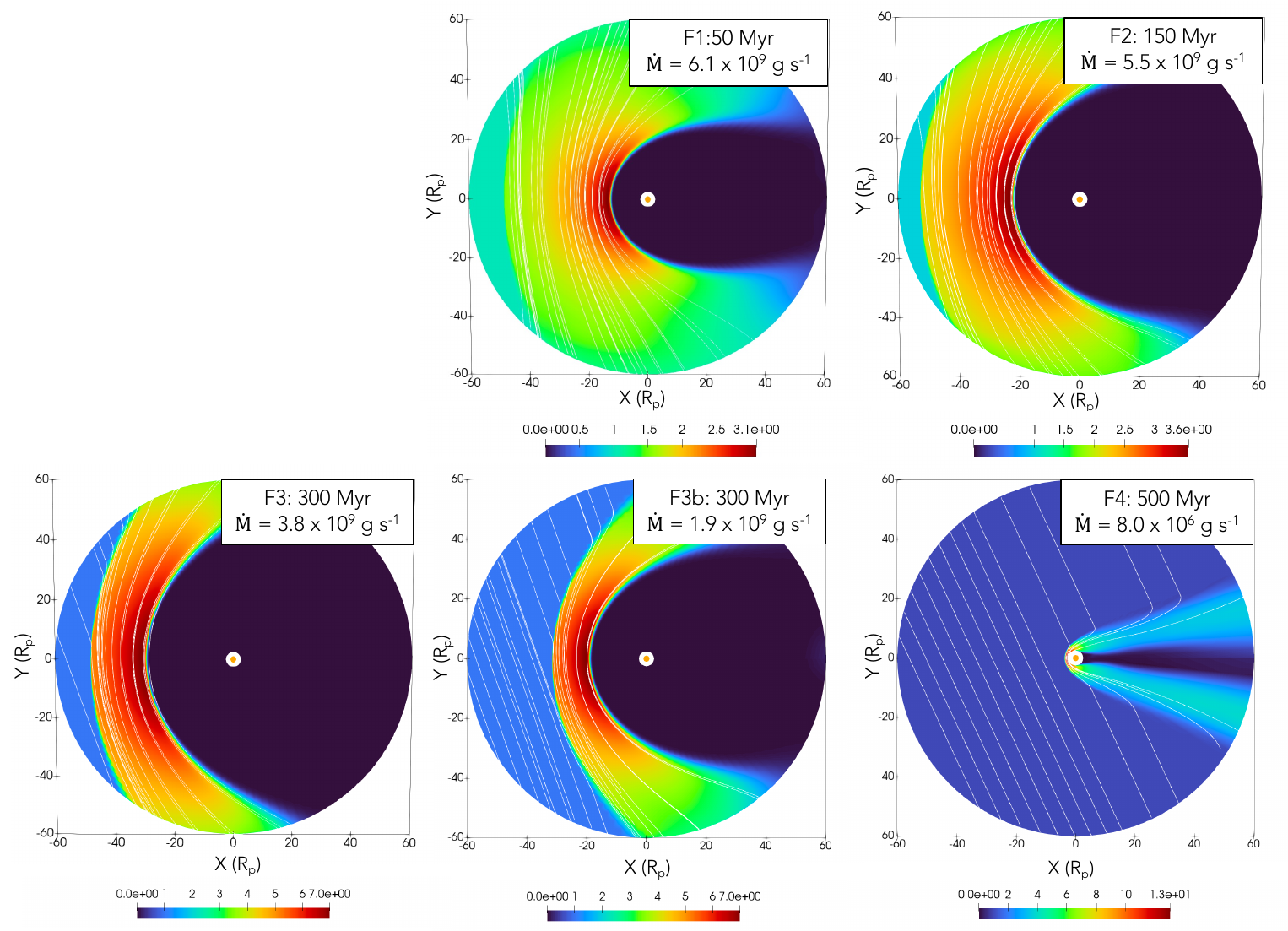}
    \caption{Magnetic field distribution for fast-rotating star regime. Magnetic field is normalized by the stellar wind magnetic field value in the age range between 50 and 500 Myr. Magnetic field streamlines are represented in thin white lines.}
    \label{fig:fast_rotator_mf}
\end{figure*}

\begin{figure*}
    \centering
    \includegraphics[scale=0.55]{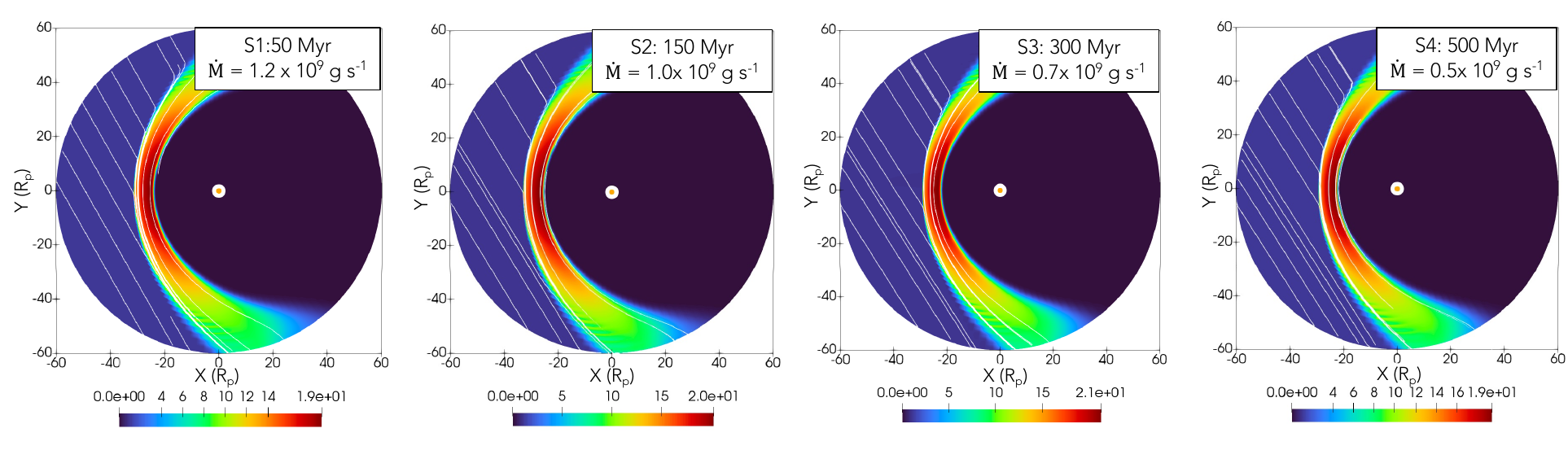}
    \caption{Same as Fig. \ref{fig:profiles_fast_rotator} but for models S1, S2, S3 and S4 corresponding to a slow-rotating star, represented respectively in columns from left to the right.}
    \label{fig:slow_rotator_mf}
\end{figure*}

Density distributions in the steady state of our simulations are shown for each model in the fast and slow-rotating stellar regimes in Fig. \ref{fig:fast_rotator_density} and Fig. \ref{fig:slow_rotator_density}, respectively. The steady state of the simulations is quickly reached at approximately $\sim$10 h, equivalent to 0.01 orbits. The first remarkable feature of the interaction of stellar winds and expanded primordial atmospheres around Earth-like planets is that the extension of the atmosphere (measured as the distance $r_{pw}$ from the planetary center limiting the region where the planetary atmosphere remains undisturbed) is significantly reduced depending on the incident stellar winds and the evolutionary state of the XUV-irradiated atmosphere. The undisturbed planetary wind can be identified in Fig. \ref{fig:fast_rotator_density} and Fig. \ref{fig:slow_rotator_density} as the region where the planetary wind flows radially from the base wind, represented in darker green. In all configurations, this region is surrounded by a higher density region (lighter green), hereafter referred as a density ``bump", a byproduct of the compression of the atmosphere due to the action of stellar winds.  

Another remarkable feature in this interaction is the formation of double sonic and alfv\'enic surfaces, beside the inner sonic surface of the planet, described in the last subsection. These surfaces are defined as the locations where the flow velocity equals the sound and Alfv\'en velocities (yellow and magenta solid lines respectively in Fig. \ref{fig:fast_rotator_density} and Fig. \ref{fig:slow_rotator_density}). The outer alfv\'enic surface constitutes the position where the stellar wind is shocked, travelling from super-fast magnetosonic (super-alfv\'enic) velocities to subalfv\'enic flow velocities, after entering into the computational domain from the left side of the grid. On the other hand, the inner sonic boundary represent the region where the planetary wind is shocked, transitioning from super- to sub-sonic velocities. Depending on the considered evolutionary state, these double surfaces are at different positions from the planet as can be observed in Fig. \ref{fig:fast_rotator_density}.

Moreover, as we considered the planetary atmosphere as a perfectly conducting fluid, modeling the interaction of the stellar winds with the uppermost ionized part of the atmospheres, the interplanetary magnetic field carried with the stellar wind is abruptly stopped when it encounters the conducting atmosphere. This leads to the accumulation of magnetic field lines above the atmospheric obstacle (see Fig. \ref{fig:fast_rotator_mf} and Fig. \ref{fig:slow_rotator_mf}), and the formation of an induced magnetosphere. An induced magnetosphere can be defined as the region inside the stellar wind shock where the magnetic pressure dominates the rest of pressure contributions \citep{2004AdSpR..33.1905L,2007Natur.450..650B,2008P&SS...56..796K}. This region is limited by two boundaries. The outer boundary, or the so called magnetopause, can be defined as the surface where the magnetic pressure equals the (dominant) dynamic pressure of the stellar wind. The second boundary, known as the ionopause, marks the point where the influence of a conductive medium, here the planetary upper atmosphere, surpasses the pressure contribution of the intensified magnetic field.

\begin{figure*}
    \centering
    \includegraphics[scale=0.4]{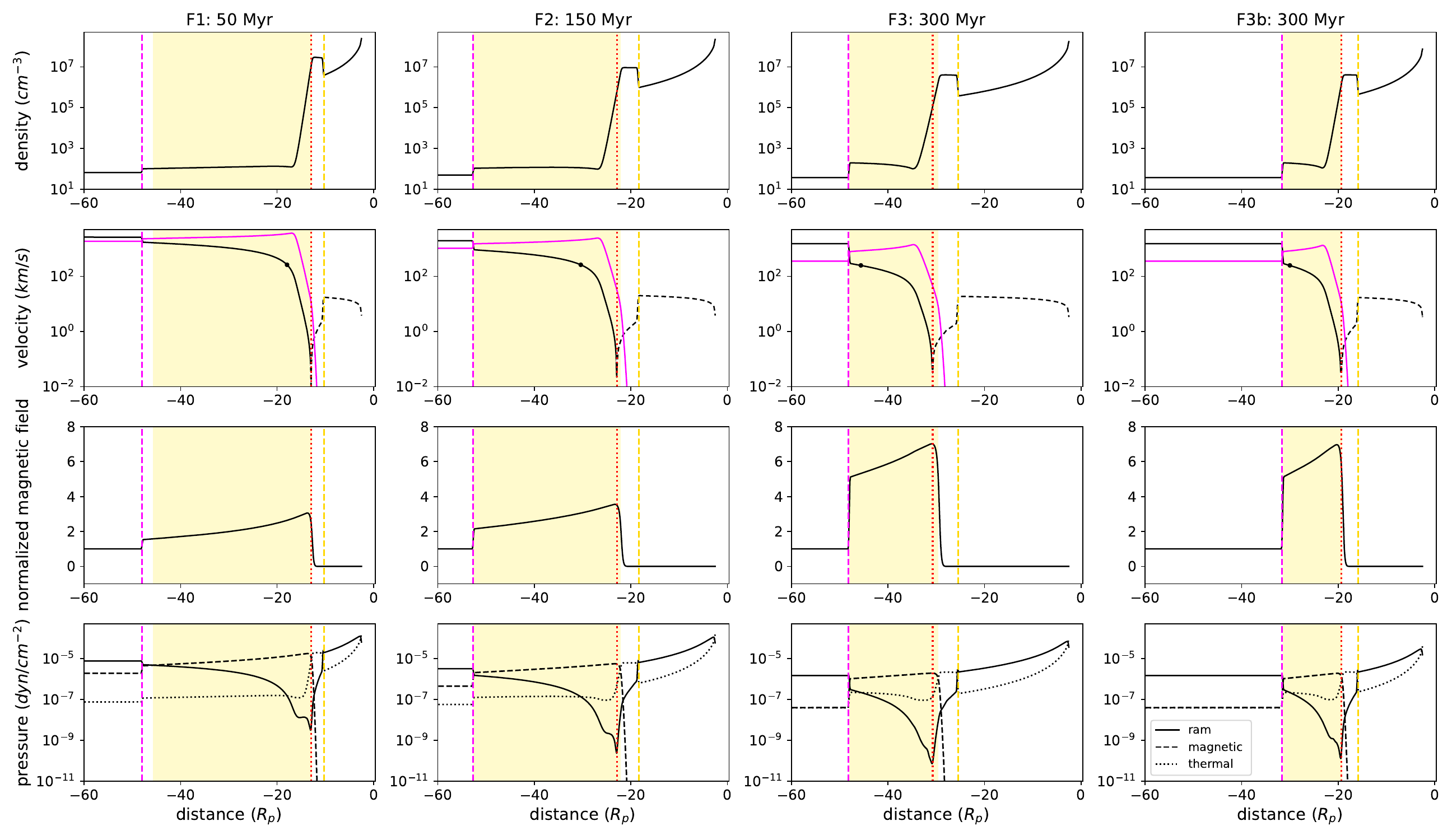}
    \caption{Profiles along the Y=0 axis of density (first row), absolute value of velocity in the X component (second row), magnetic field normalized by the stellar wind interplanetary magnetic field (third row), and pressure profiles accounting for dynamic (ram, solid line), thermal (dotted line) and magnetic (dashed line) contributions, for each single model (F1 to F3b, from left to right columns). Vertical magenta and yellow dashed lines represent the location of the stellar and planetary wind shocks, respectively. Vertical dotted red line represents the point $r_X$ where velocity experiences a sign reversal. In the second row, the flow velocity is represented by a black line, where the solid part of the profiles corresponds to positive velocities (stellar wind contribution) and dashed part corresponds to negative velocities (planetary wind contribution). The Alfv\'en velocity is represented by the solid magenta line. The point where the stellar wind becomes subsonic is represented by the black dot in second row. The dashed yellow region corresponds to the extent of the induced magnetosphere.}
    \label{fig:profiles_fast_rotator}
\end{figure*}

\begin{figure*}
    \centering
    \includegraphics[scale=0.4]{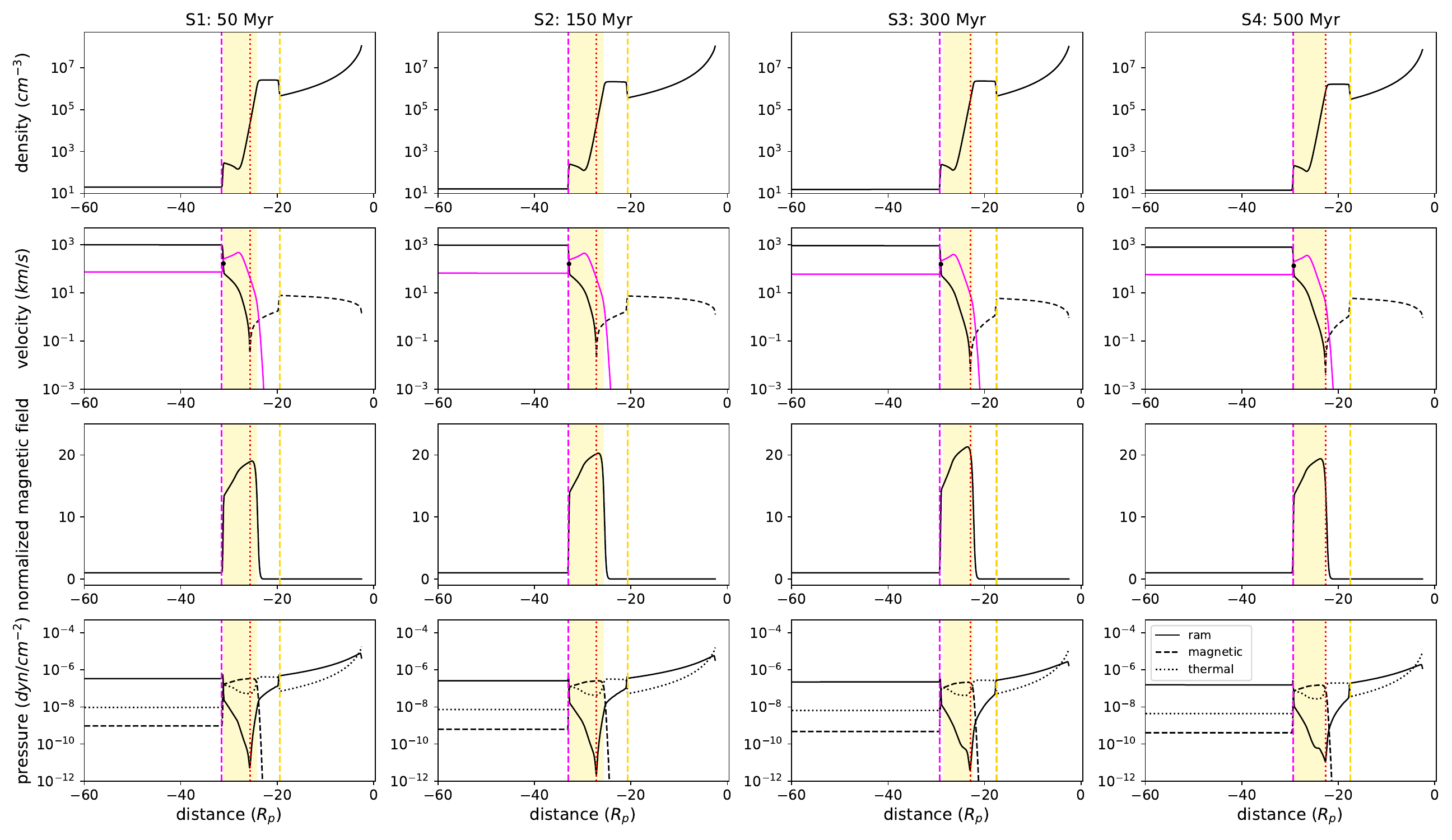}
    \caption{Same as Fig. \ref{fig:profiles_fast_rotator} but for models S1, S2, S3 and S4 corresponding to a slow-rotating star, represented respectively in columns from left to right.}
    \label{fig:profiles_slow_rotator}
\end{figure*}

In Fig. \ref{fig:profiles_fast_rotator} and Fig. \ref{fig:profiles_slow_rotator} the profiles of density, absolute value of the X velocity component, magnetic field normalized by the local interplanetary field value and pressure contributions along the substellar line ($X<0,Y=0$) are represented for fast and slow-rotating regimes, respectively. Profiles for model F4 are not represented here, as the interaction region is too small for comparison in that case. Similar profiles are obtained in all configurations, sharing a common interaction and structure formation. The stellar wind enters undisturbed from the left side boundary (X = - 60 R$_p$) and is shocked at a distance $r_{sw}$, represented by a vertical magenta dashed line, and coincident with the outer alfv\'enic surface. The main pressure contribution in the stellar wind side of the shock is the ram pressure, defined as $p_{ram} = \rho v^2$. Across the shock, density and tangential (oriented in the Y direction) magnetic field are increased, while the flow velocity is decreased. This increase/decrease of the fundamental variables across the shock, usually referred as the shock strength, is dependent on the compression of the magnetosonic wave, measured through the fast magnetosonic Mach number $M_{fm}$ of the wind, with lower $M_{fm}$ (low-compression) leading to a fainter increase (decrease) in density and magnetic field (velocity). The strength of the shock, measured as the ratio between the upstream (before the shock) and downstream (after the shock) values of density, velocity and magnetic field is 1.5 for model F1 ($M_{fm}$ = 1.4). For model F2 ($M_{fm}$ = 1.8), the ratio is increased to 2.2. For models F3 and F3b ($M_{fm} = 3.5$), the strength of the shock is increased to 5.2. For planets orbiting a slow-rotating star, the shock strength is approximately constant with values around 14, in agreement with nearly constant considered $M_{fm}$ numbers ($M_{fm}\sim 5.5$).

After the stellar wind shock, the interaction becomes subalfv\'enic. For more compressing stellar winds (higher $M_{fm}$), the stellar wind becomes also subsonic just after the shock. For low $M_{fm}$, as considered in models F1 and F2, the stellar wind becomes subsonic at further distances from the stellar wind shock (see second row in Fig. \ref{fig:profiles_fast_rotator}).

Both planetary and stellar wind meet at a distance $r_X$ (red dotted vertical line in Fig. \ref{fig:profiles_fast_rotator} and Fig. \ref{fig:profiles_slow_rotator}), where the velocity of the flow experiences a sign reversal (from positive inflow stellar wind velocities to negative outflow planetary wind velocities in the X component), corresponding to a minimum in velocity magnitude and ram pressure (fourth row in Fig. \ref{fig:profiles_fast_rotator} and Fig. \ref{fig:profiles_slow_rotator}). 

The formation of the induced magnetosphere in all configurations can be observed in the magnetic field profiles (see third row in Fig. \ref{fig:profiles_fast_rotator} and Fig. \ref{fig:profiles_slow_rotator}). The magnetic field is increased from the stellar wind shock, being accumulated in front of the ionized upper atmosphere, and reaching a maximum close to the meet point $r_X$. This magnetic field enhancement results in an increase in the general magnetic pressure of the system, resulting in the formation of an induced magnetosphere. In Fig. \ref{fig:profiles_fast_rotator}, the extent of the induced magnetosphere is represented by a shaded yellow region.

The maximum value of the accumulated magnetic field is again dependent on the Mach numbers of the incident stellar wind, being increased for high $M_{fm}$ numbers. For planets orbiting a fast-rotating star, the maximum of the magnetic field is increased up to 3 times the interplanetary magnetic field value at 50 Myr ($M_{fm}=1.4$), slightly increasing this value to 3.6 at 150 Myr ($M_{fm}=1.8$), and increasing again up to 7 times the interplanetary magnetic field for models F3 and F3b ($M_{fm}=3.5$). At 500 Myr, the wind is characterized with a higher $M_{fm}$ of 5.1, leading to an increase of the magnetic field about 13 times the interplanetary magnetic field value. For stars in the slow-rotating regime, stellar winds are characterized with higher magnetosonic values about $M_{fm}$ = 5.5, leading to an amplification of the magnetic field inside the induced magnetosphere around 20 times the interplanetary magnetic field, remaining approximately constant around this value. 

From the planetary wind base, the planetary outflow travels from subsonic to supersonic velocities, remaining unaltered until it is shocked at a distance $r_{pw}$ (represented by a dashed yellow vertical line in Fig. \ref{fig:profiles_fast_rotator} and Fig. \ref{fig:profiles_slow_rotator}), being decelerated again to subsonic velocities. The main pressure contribution in the planetary wind side is the dynamic pressure, even though low characteristic flow velocities are considered, as described in the last subsection. The distance $r_{pw}$ defines the extension of the planetary atmosphere, below which the planetary wind remains undisturbed. From $r_{pw}$, the velocity is dropped to nearly zero values and the density abruptly increased, forming a density ``bump" as a result of the compression of the incident stellar wind and low velocities of both winds close to the interaction region (see first row in Fig. \ref{fig:profiles_fast_rotator}). This density ``bump" increases the density up to approximately 10 times the value of atmospheric density just before the shock in all stellar wind-planet configurations.

\subsection{Evolution of escaping atmospheres: fast-rotating stars}
The evolution of stellar activity in the age range between 50 and 500 Myr for rapidly rotating stars is characterized by a first stellar saturation period ending at 260 Myr (encompassing models F1 and F2) characterized by an enhanced XUV stellar emission ($F_{XUV}\sim$ 3400 erg s$^{-1}$ cm$^{-2}$) and strong stellar winds. The escaping atmospheres are extremely inflated due to the enhanced XUV activity of the star, resulting in high densities and temperatures at large distances from the planet. From 260 Myr (models F3, F3b and F4), the star become unsaturated, with a decreased XUV emission to approximately 2268 erg s$^{-1}$ cm$^{-2}$ at 300 Myr, dropping to 542 erg s$^{-1}$ cm$^{-2}$ at 500 Myr. Stellar winds are also fainter due to the decreased stellar magnetic activity, as described in Section \ref{sec:paperIII_winds}.

Following the results of our MHD numerical simulations, the location of the previously described boundaries and structures is found to be dependent on the pressure balance between the incoming stellar wind and the pressure exerted by the dynamical planetary wind. Moreover, the formation of the induced magnetosphere contributes with additional magnetic pressure pushing the atmospheric envelope towards the planet, but also pushing outwards the incident stellar wind, leading to the stellar wind shock to move further away from the planet. As described in \citet{2023MNRAS.525..286C}, the stellar wind bow shock position when an induced magnetosphere is formed is found to be sensitive to the Alfv\'en Mach number of the stellar wind, with more extended shocks for low Alfv\'en Mach numbers. 

At 50 Myr the atmosphere is pushed towards the planetary surface by the stellar wind, leading to a reduced atmospheric extent of 10.3 R$_p$. At this evolutionary stage, the ram pressure of the stellar wind is $7.6\times10^{-6}$ dyn cm$^{-2}$, while the ram pressure of the planetary wind (measured at a close distance to the wind base at 3 R$_p$) is $9.4\times10^{-5}$, as a result of the high density atmospheric levels. The stellar wind bow shock is found at 48.0 R$_p$ from the planetary center. The point $r_X$ where both winds meet is located slightly above the undisturbed planetary wind at 12.3 R$_p$. Moreover, the planetary wind flow is also suppressed by the stellar wind behind the planet (X$>0$) at 30.2 R$_p$, in the so called planetary wind tail region, as can be seen in the center upper panel in Fig. \ref{fig:fast_rotator_density}. 

At 150 Myr, the stellar wind strength is decreased due to a reduction in the star's rotation rate, resulting in less accelerated winds, decreasing the characteristic ram pressure of the stellar wind to $3.2\times10^{-6}$ dyn cm$^{-2}$ (2.4 times lower than at 50 Myr). On the other hand, the atmosphere is still hot and inflated due to the constant XUV emission of the star. However, the density of the atmosphere is slightly reduced due to faint atmospheric losses, leading to a decrease in the atmospheric ram pressure of only 1.1 times lower than at 50 Myr. This results in an increase of the atmospheric extension, moving further away from the planet up to 18.2 R$_p$. The stellar bow shock is located at larger distances from the planet's center at 52.6 R$_p$, due to the pressure exerted by the atmosphere. Even though at this evolutionary stage the pressure exerted by the induced magnetosphere will move further away the shock from the planet, as a consequence of the higher $M_A$ of the stellar wind, the pressure exerted by the atmosphere is dominant in the process. The meet point $r_X$ is also located at further distances from the planet, at 22.6 planetary radii. At this evolutionary stage, the planetary tail is not affected by the action of the stellar wind, where planetary wind flows undisturbed.

At 300 Myr (models F3 and F3b), the star abandons the saturated regime, resulting in a decrease of the strength of the stellar wind.  At the same time, the atmosphere is less inflated due to the decrease of the XUV radiation and the thermal losses experience up to this evolutionary stage. In model F3, considering a characteristic reduced atmospheric mass fraction of $1\times10^{-3}$, the extension of the planetary atmosphere is however increased up to 25.3 R$_p$. The ram pressure of the stellar wind has decreased 2.1 times with respect to 150 Myr values, while the atmosphere has only decreased 1.4 times its ram pressure, resulting in a more expanded atmosphere. As a consequence, the point $r_X$ is located at further distances as well at 30.3 R$_p$. The bow shock of the stellar wind is located at 48.2 planetary radii, closer to the planet in comparison to model F2. This can be attributed due to the higher $M_A$ number ($M_A$ = 4.2) of the stellar wind in comparison to model F2 ($M_A$ = 1.8), resulting in a closer stellar wind shock, as the magnetic pressure exerted here by the induced magnetosphere is much lower than the injected ram pressure of the stellar wind. As in the 150 Myr case, the planetary tail is not perturbed as a consequence of the stellar wind interaction.

Approximately at the same stellar age, the planet undergoes strong atmospheric losses (\citetalias{2015ApJ...815L..12J}). In model F3b, we reduced the mass-loss rate and density of the planetary atmosphere, keeping the same stellar wind conditions. In this way, the ram pressure exerted by the planet is decreased 2.3 times with respect to the ram pressure defined in model F3. The atmosphere extension is reduced down to 15.8 R$_p$. The distance $r_X$ is as well reduced down to 19.3 radii from the planet, and the stellar wind bow shock position is found closer to the planet at 31.6 planetary radii, due the reduced pressure exerted by the planetary wind. As described in model F1, the stellar wind also affects the tail of the planet, reducing its extension approximately up to 47.1 planetary radii. 

For comparison, we run an additional model (F4) at 500 Myr, when the atmosphere is supposed to be completely photoevaporated. The interaction of the stellar wind at this considered evolutionary stage and the almost absent atmosphere results in a drastic different interaction. The stellar wind is shocked at very close distances from the base wind. As in the other models, the interplanetary magnetic field is draped around the planet, forming a reduced induced magnetosphere. However, the inner limit of the induced magnetosphere is located at the atmosphere base, as no atmospheric gas can push outwards the stellar wind. 

The extension of the induced magnetosphere is reduced with stellar age. The induced magnetosphere outer boundary (magnetopause), where the magnetic pressure is higher than the rest of pressure contributions, is located just before the stellar wind shock. The inner boundary of this magnetic field dominated region is located approximately at the point where both stellar and planetary winds meet. This structure is dependent therefore on the pressure exerted by the planetary wind, moving outwards the meet point $r_X$, and the location of the stellar wind bow shock, controlled by the pressure exerted both by the planetary and stellar winds, and the extra pressure exerted by induced magnetosphere. At 50 Myr, the induced atmosphere extends about 33 planetary radii, being reduced up to 13 planetary radii in model F3b (see shaded yellow region in Fig. \ref{fig:profiles_fast_rotator}). At 500 Myr, the induced magnetosphere is compressed with an extension of 1.1 R$_p$ .

\subsection{Evolution of escaping atmospheres: slow-rotating stars}

For an evolving planet orbiting a slowly rotating star, the received XUV fluxes and stellar winds of these stars are significantly weaker compared to those described for fast-rotating stars. Furthermore, both the stellar XUV radiation and stellar winds keep nearly constant values. The XUV fluxes range between 224 and 95 erg s$^{-1}$ cm$^{-2}$, leading to nearly constant atmospheric mass-loss rates and density profiles. Additionally, the winds from these stars gradually weaken over time, albeit not markedly, with accelerated winds reaching up to 1000 km/s at 50 Myr, decreasing to 800 km/s at 500 Myr.

As expected, escaping atmospheres undergo minimal changes during the considered period. At 50 Myr, the planetary atmosphere extends up to 19.6 planetary radii. In this evolutionary scenario, the dynamic pressure exerted by the stellar wind is $3.4\times 10^{-7}$ dyn cm$^{-2}$, an order of magnitude lower than wind pressures for a fast rotator at the same epoch. The stellar wind and the escaping atmosphere meet at $r_X=25.6$ R$_p$ from the center of the planet. The stellar wind bow shock is found at 31.4 R$_p$.

By 150 Myr, the atmosphere extent is slightly increased to 20.7 planetary radii. The dynamic pressure exerted by the stellar wind merely decreases to 1.3 times the pressure at 50 Myr. Additionally, the atmospheric density is reduced by a factor 1.1; the atmosphere is then slightly pushed outwards. However, the interaction remains nearly identical, with similar locations for the stellar wind bow shock, found at 32.8 R$_p$ and the meet point $r_X$ found at 27.1 R$_p$.

At 300 Myr, the atmospheric extent slightly decreases to about 17.5 planetary radii. Similarly, the bow shock also moves slightly closer to the planet at 29.3 R$_p$. In comparison to the conditions defined at 150 Myr, the stellar wind dynamic pressure has only decreased by 1.5 times, while the atmospheric density has reduced to 1.8 times, resulting in this reduction in atmospheric extent. Both winds meet at $r_X=22.9$ R$_p$.

At 500 Myr, the atmosphere remains entirely stable compared to its characteristics described at 300 Myr. The atmospheric extent, the position of the stellar wind bow shock, and the critical distance $r_X$ remain unchanged.

The planetary tail is not perturbed by the action of the stellar wind considering a slow-rotating star. 

The extension of the induced magnetosphere is drastically reduced in comparison to the fast-rotating regime. In the slow-rotating regime, the extension of this structure remains approximately unchanged, with an extension of 7 planetary radii in models S1 and S2, and 6 planetary radii in models S3 and S4. 

\begin{figure*}
    \centering
    \includegraphics[scale=0.5]{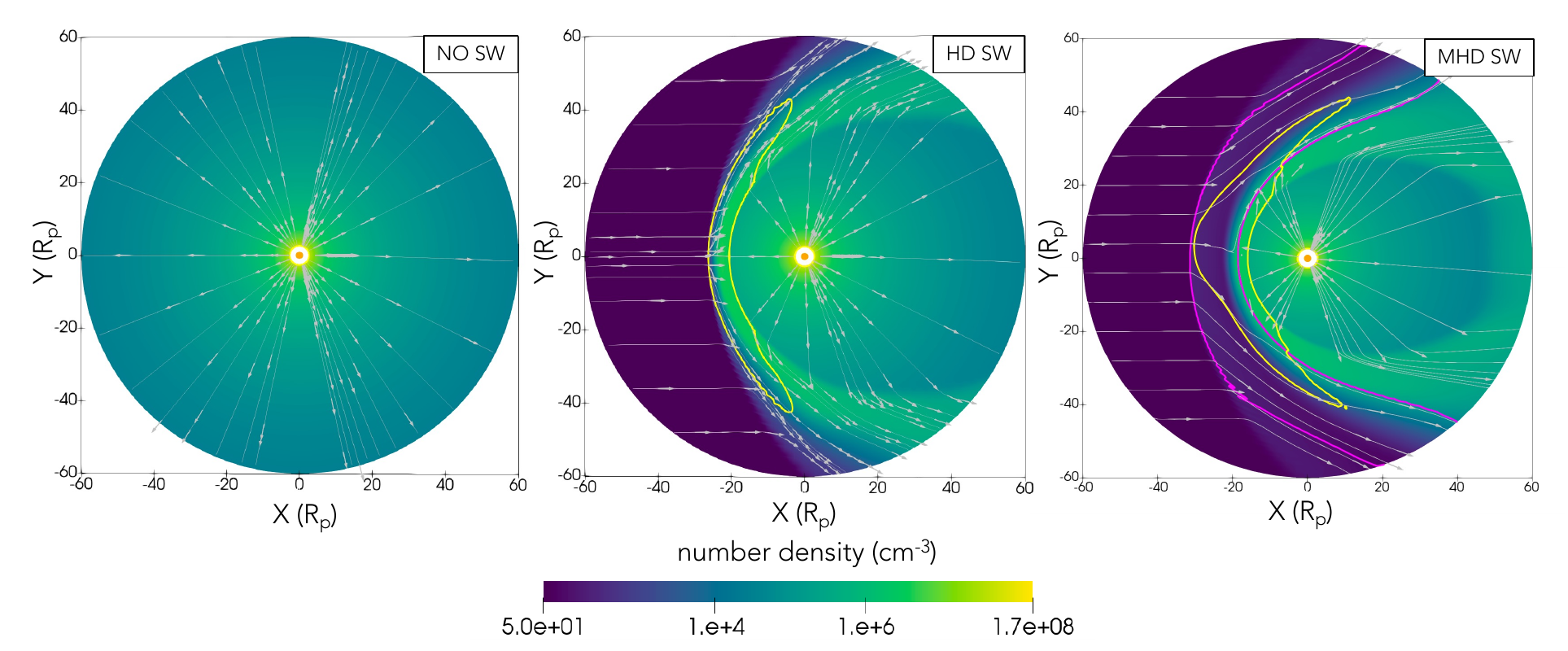}
    \caption{Number density distribution in the steady state for an atmosphere in the absence of stellar wind action (left), the hydrodynamic interaction (center), and the MHD interaction (right) for the stellar wind-planet   F3b case.}
    \label{fig:HD_MHD}
\end{figure*}

\subsection{Atmospheric mass-loss rates}
In all addressed models, the interaction between the stellar wind and the escaping atmosphere occurs beyond the planet sonic surface, implying that the planetary mass-loss rate is not reduced from the wind base, as described in \citet{2016ApJ...820....3C,2020MNRAS.494.2417V}. Instead, due to the interaction with the stellar wind, the atmosphere experiences a significant compression, resulting in a density ``bump" that is dragged outwards by the stellar wind at very low velocities, particularly in the flanks region, thereby causing an increase in atmospheric mass-loss.

To estimate the average atmospheric mass-loss, we integrated the mass flux across spheres centered on the planet of varying radii within the region corresponding to the density bump, as:
\begin{equation}
    \dot{M} = \oint{\rho \mathbf{v}} \cdot dS
\end{equation}
where the integral is performed across spheres of varying radii. We then compared the obtained values with the mass-loss rates values inside the atmospheric unaltered region (see Table \ref{tab:atm_param}).

The atmospheric mass-loss rate is increased by 1\% in model F1 (fast-rotating star and expanded atmosphere). The mass-loss is higher in models F2 and F3, increasing up to 4\%. For model F3b, the mass loss is lower, being increased by only 1 per cent.

In the low-rotation star regime, similar increases in mass-loss are observed, with a 4\% increase for models S1 and S2, and a 2\% increase for models S3 and S4.

The differences observed in the increased planet mass-loss from one model to another is likely to lie in the extent of the interaction region, increasing the area where the density is increased and dragged out by the action of the stellar winds. 

Nevertheless, the values obtained for the increase in mass-loss are much lower compared to the mass-losses triggered by the XUV stellar radiation.

\subsection{Importance of interplanetary magnetic fields}
The interplanetary magnetic field carried by the stellar wind plays a fundamental role in the star-planet interaction, particularly in the case of unmagnetized planets, as highlighted in our previous findings. The magnetic field not only triggers the formation of new structures, such as the induced magnetosphere, but also alters the extension of other structures, as observed in the position variation of the stellar wind bow shock.

To assess the influence of the interplanetary magnetic field on atmosphere extension, additional simulations were conducted under identical parameters, but suppressing the interplanetary magnetic field to study a pure hydrodynamic (HD) interaction. Figure \ref{fig:HD_MHD} shows the density distribution in the steady state for an hydrodynamic escaping atmosphere in the absence of stellar wind action (left), the HD interaction (center), and the magnetohydrodynamic interaction (right) for the previously described case F3b.

In the absence of stellar winds, the planetary atmosphere extends farther away, reaching 20.3 planetary radii in the HD scenario, compared to 15.8 R$_p$ in the MHD case. Additionally, the tail extension remains undisturbed in the hydrodynamic scenario, while it undergoes reduction in the magnetohydrodynamic approximation.

Furthermore, the position of the stellar wind's bow shock is also altered, located at 27.0 planetary radii in the HD case, contrasting with 31.6 radii in the MHD scenario.

Consequently, the morphology of the double sonic surface around the planet is also modified, as a consequence of the extent of the interaction region limited by the stellar and planetary wind bow shocks, leading to larger sonic surfaces in the MHD case. 

These significant differences are spread consistently across all the planet-wind configurations described earlier, showing a widespread reduction in atmosphere extension when accounting for the influence of the interplanetary magnetic field, as a consequence of the additional magnetic pressure exerted on the planet by the induced magnetosphere.

\section{Discussion and Conclusions}
\label{sec:discussion_paperIII}
\begin{figure}
    \centering
    \includegraphics[scale=0.5]{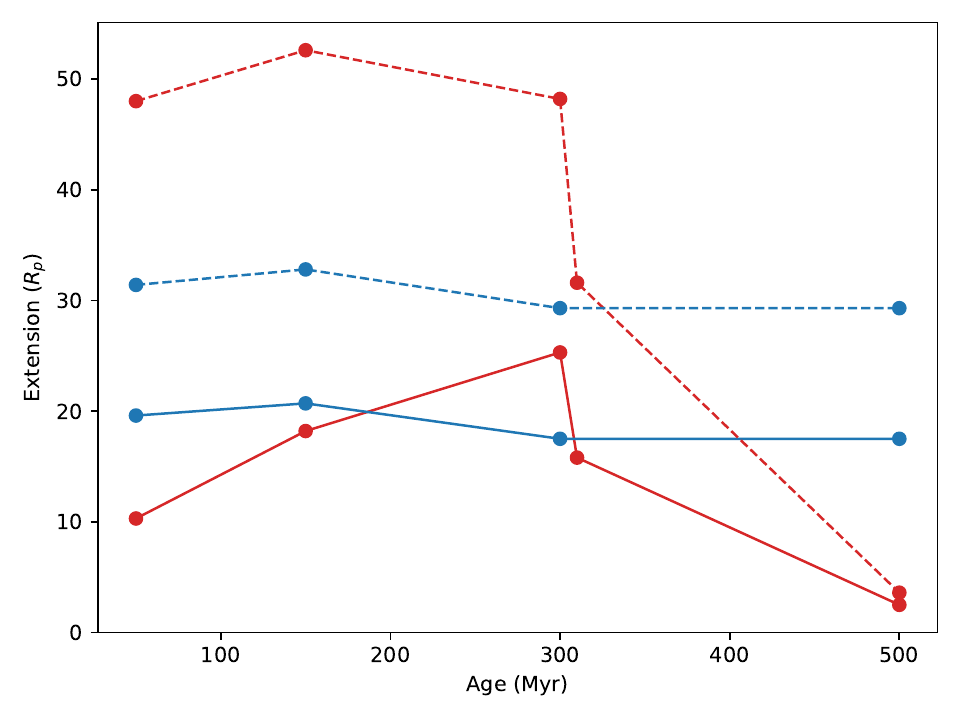}
    \caption{Atmospheric extensions (solid lines) and stellar wind bow shock positions (dashed lines) measured from the planetary center in units of planetary radius, considering fast (red) and slow (blue) rotating stars, as a function of age.}
    \label{fig:extensions}
\end{figure}

\begin{table*}
\label{tab:IM}
\caption{Comparison of induced magnetosphere (IM) extensions and maximum of the magnetic field inside the induced magnetosphere compared to the local interplanetary magnetic field values for all studied cases.}
\begin{tabular}{cccc}
\hline
\hline
\multicolumn{4}{c}{Induced magnetospheres: fast rotator}                                  \\ \hline\hline
model & Age (Myr) & IM extension (R$_p$) & Max. MF \\
\hline

F1    & 50        & 33 & 3.0 \\
F2    & 150       & 30 & 3.6 \\
F3    & 300       & 17 & 7.0 \\
F3b    & 300       & 13 &  7.0 \\
F4    & 500      &  1.1  &  13.0 \\ 
\hline
\hline
\multicolumn{4}{c}{Induced magnetospheres: slow rotator}                                  \\ \hline \hline
model & Age (Myr) & IM extension (R$_p$) & Max. MF \\
\hline

F1    & 50        & 7 & 19.0 \\
F2    & 150       & 7 & 20.2 \\
F3    & 300       & 6 & 21.3 \\
F4    & 500      &  6  &    19.5   \\ 
\hline
\hline
\end{tabular}
\end{table*}
In this paper, we evaluate the impact of magnetized stellar winds on hydrodynamic primordial atmospheres surrounding unmagnetized Earth-like planets orbiting at 1.0 au in the age range between 50 and 500 Myr. We explore two distinct regimes based on stellar activity, examining the action of stellar winds originating from fast and slow-rotating stars.

For evaporating planets orbiting fast-rotating stars, planets exhibit expanded and hot atmospheres due to the enhanced XUV emission of the star, especially inside the stellar saturated regime. Moreover, this enhanced magnetic stellar activity results also in fast and hot stellar winds. The joint evolution of both XUV-exposed atmospheres and stellar winds results in a significant atmospheric confinement at early ages around 50 Myr, reducing the atmospheric extent to approximately 10 planetary radii. The planetary atmosphere becomes less confined from 150 Myr in comparison to early ages, reaching a maximum extension of more than 25 planetary radii at 300 Myr, even though part of the atmosphere has been lost by means of hydrodynamic thermal escape, and atmospheres are cooler as a consequence of the decrease in the XUV radiation levels.  Our simulations reveal that in a first approximation, the stable morphology of these escaping atmospheres is governed by the pressure balance between both stellar and planetary wind plasmas. In the first 300 Myr, the evolution is governed by the evolution of stellar winds. From 300 Myr, as a consequence of massive photoevaporation of the planetary atmosphere, resulting in a higher confinement by the stellar wind, the atmosphere is reduced again to 15.8 R$_p$ (see Fig. \ref{fig:extensions}).

For fast-rotating stars, we also explored the influence of this plasma for saturated and unsaturated stars. However, no significant differences have been observed between saturated stars at 150 Myr and unsaturated stars at 300 Myr. This is because at 300 Myr, the stars have just abandoned the saturated regime, approximately at 260 Myr, and stellar wind parameters do not show large modifications. At 500 Myr, the interaction has been dramatically changed, as both the atmosphere and the stellar wind strength have been reduced due to the decrease of the stellar activity, resulting in a close planet interaction.

The significant reduction specially at early ages and for lower-mass escaping atmospheres could have relevant consequences in terms of detectability, leading to a decrease in Lyman-$\alpha$ absorption during the planet's transit, assuming that the neutral component of the atmosphere will follow the ionized component via collisions. In addition, at the fixed orbital distance set at 1.0 au, no significant asymmetries are observed in the morphology of these atmospheres, because the pressure induced by the winds exceeds that caused by the planet's orbital motion, as proven to be important for close-in exoplanets \citep[e.g.][]{2015A&A...578A...6M,2021MNRAS.500.3382C,2022MNRAS.510.3039K,2021MNRAS.501.4383V}.

For Earth-like planets orbiting slow-rotating stars, no significant changes have been reported in the 50-500 Myr period (see blue lines in Fig. \ref{fig:extensions}). This period is characterized by moderate and stable stellar activity, where planetary atmospheres remain almost unchanged, experiencing only minor mass-losses. Additionally, the stability in stellar activity results in moderate winds, exhibiting few alterations between the 50 to 500 Myr period. Atmospheres during this time span extend up to 20 planetary radii, with a slightly reduced extension after 300 Myr, yet without significant changes in their distribution.

The effects of interplanetary magnetic fields carried by stellar winds have been assessed, primarily resulting in the formation of induced magnetospheres around these planets. The extension of this region, characterized by sub-Alfv\'enic speeds and a predominant magnetic field influence, universally decreases with age, diminishing from 33 to 1.1 from 50 to 500 Myr for rapidly rotating stars, whereas this region is significantly smaller (between 6 and 7 planetary radii) for slowly rotating stars. Moreover, the maximum magnetic field intensity achieved inside the induced magnetosphere are dependent on the magnetic properties of the stellar winds, with higher magnetic field intensities for more compressive (higher $M_{mf}$) winds. Main results obtained in this work are shown in Table \ref{tab:IM}. 

Moreover, we compared the MHD interaction performed in this work with the HD interaction, keeping the same values for atmospheric and stellar wind ram pressures but neglecting the presence of magnetic fields. Incorporating magnetic fields into the stellar wind-planet interaction leads to further confinement of the planetary atmosphere due to the added magnetic pressure. Additionally, it results in stellar wind bow shocks positioned farther away, influenced by the induced magnetospheres. This accentuates the significance of acknowledging the substantial role of magnetic fields in these interactions, and the limitations of studying stellar wind-planet interactions under the hydrodynamic approach. 

In the study by \citet{2022ApJ...934..189C}, it is also demonstrated that incorporating the interplanetary magnetic field into planet-stellar wind  simulations is crucial. In the case of the short orbit planet AU Microscopii b, this inclusion increases the total pressure in the vicinity of the planet, thus altering the properties of the planetary wind, primarily accelerated by a pressure gradient. Additionally, our study shows that in the case of more distant planets, where the density and magnetic field of the stellar wind are lower, even though the planetary wind structure is not changed from the base, the interplanetary magnetic field plays a fundamental role in the formation of structures around the planet, contributing to additional confinement of the planetary atmosphere through the action of magnetic pressure.

This study assumes the atmosphere as a fully ionized plasma. The magnetic field is drastically stopped by the presence of the conductive atmosphere, leading the inner atmosphere unperturbed. On the flanks, the ionized material from the outer part of the atmosphere is slowly dragged by the stellar wind, following the magnetic field lines. The consequent atmospheric drag by the stellar wind results in additional mass-loss; however,  mass-loss rates are insignificant compared to losses induced by the stellar low-wavelength radiation.

In summary, this study highlights the extent to which stellar wind interaction cannot be disregarded in the early evolution of primordial planetary atmospheres, as it stands as a pivotal factor shaping the morphology of these gaseous envelopes and their detectability. Nevertheless, accounting to the mass-loss rated induced by the action of stellar winds, we can conclude that the overall losses of these atmospheres remains primarily governed by the influence of the star's photoevaporative radiation.

\section*{Acknowledgements}
This work has been partially financed by the Ministry of Science and Innovation through grants: ESP2017-87813-R and PID2020-116726RB-I00, US DOE under grant DE-FG02-04ER54742 and the project 2019-T1/AMB-13648 founded by the Comunidad de Madrid. 

\section*{Data Availability}
The data underlying this article will be shared on reasonable request to the corresponding author.



\bibliographystyle{mnras}
\bibliography{mybibliography} 




\appendix

\section{Comparison between Divergence Cleaning and Eight Waves methods for controlling the magnetic field divergence in the planet-stellar wind interaction}
\label{ap1}
In this study, we performed numerical simulations of magnetized stellar winds and planetary atmospheres using two different approximations for controlling magnetic field divergence: the divergence cleaning (DC) and the eight waves (8W) approximations, that are currently implemented in the \textsc{pluto} code. The use of the DC method is widespread in numerical MHD simulations, including the planet-stellar wind interaction \citep[e.g.]{2023MNRAS.525.4008V,2015A&A...578A...6M} in contrast to the use of the 8W method, as the latter keeps the solenoidal condition for the magnetic field only at the truncation level and not to machine accuracy. We refer to \citet{1994arsm.rept.....P} and \citet{2002JCoPh.175..645D} for a fully description of the DC and 8W methods respectively. 

\begin{figure}
    \centering
    \includegraphics[scale=0.52]{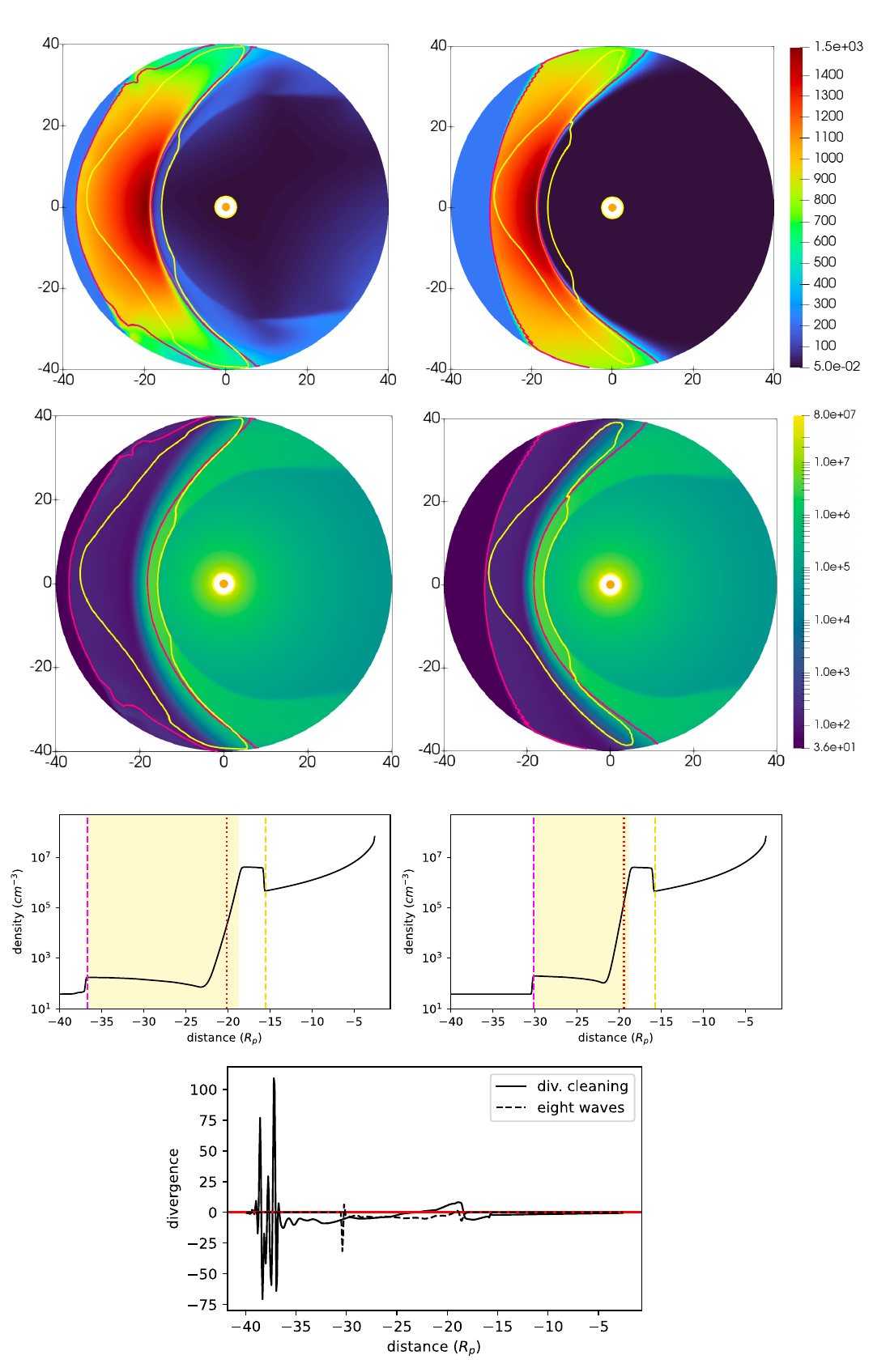}
    \caption{First, second and third rows show the magnetic field distribution, number density distribution, and density profiles along the substellar line (Y=0), using the DC (first column) and 8W (right column) magnetic field divergence control methods. The forth row shows the divergence of the magnetic field along the substellar line for the DC method (solid line) and the 8W method (dashed line). The red horizontal line marks the zero threshold.}
    \label{fig:dc_8W}
\end{figure}

Here, we show the interaction under both approximations for same atmospheric and stellar wind conditions following the adopted parameters for F3b case. In Fig. \ref{fig:dc_8W}, the magnetic field (first row), number density (second row) and density profiles along the substellar line (third row) are shown using the DC method (left column) and the 8W method (right column). The density structure of the interaction is preserved under both approximations, with no significant differences especially in the stellar wind-atmosphere interaction region. However, the stellar wind bow shock position is altered, being further away (around 37 planetary radii) when considering the DC method, in comparison to the 8W method. At the bow shock position, strong magnetic turbulence is found when adopting the divergence cleaning method, specially at the outer boundary of the simulation domain. Moreover, we calculate the magnetic field divergence (see bottom panel of \ref{fig:dc_8W}), showing that this condition is not well preserved in the shock region for the DC method, resulting in magnetic field turbulence that modifies the extension of the bow shock. The 8W method profile shows a near zero divergence in the whole domain, so this method is the preferred approximation adopted in this work. 


\bsp	
\label{lastpage}
\end{document}